\documentclass[preprintnumbers,aps,prd,floatfix,groupedaddress,nofootinbib,twocolumn,superscriptaddress]{revtex4-1}

\usepackage{amsmath,amssymb,bm,color,dcolumn,float,mathrsfs,tabularx,graphicx}
\usepackage{hyperref}
\hypersetup{colorlinks=true,linkcolor=blue,citecolor=blue,urlcolor=blue}
\usepackage{Macro} 

\newcommand{\bfk}{\mbox{\boldmath$k$}}

\begin{document}

\title{CMB lensing  bi-spectrum: assessing analytical predictions against full-sky lensing simulations}

\author{Toshiya Namikawa}
\affiliation{Leung Center for Cosmology and Particle Astrophysics, National Taiwan University, Taipei, 10617, Taiwan}
\author{Benjamin Bose}
\affiliation{Yukawa Institute for Theoretical Physics, Kyoto University, Kyoto 606-8502, Japan}
\affiliation{Departement de Physique Theorique, Universite de Geneve, 24 quai Ernest Ansermet, 1211 Geneve 4, Switzerland}
\author{Fran\c{c}ois R. Bouchet}
\affiliation{Institut d'Astrophysique de Paris, UMR 7095, Sorbonne Universit\'e \& CNRS, 75014 Paris, France}
\author{Ryuichi Takahashi}
\affiliation{Faculty of Science and Technology, Hirosaki University, 3 Bunkyo-cho, Hirosaki, Aomori 036-8561, Japan}
\author{Atsushi Taruya}
\affiliation{Center for Gravitational Physics, Yukawa Institute for Theoretical Physics, Kyoto University, Kyoto 606-8502, Japan}
\affiliation{Kavli Institute for the Physics and Mathematics of the Universe, The University of Tokyo Institutes for Advanced Study (UTIAS), The University of Tokyo, Chiba 277-8583, Japan}
\date{\today}

\begin{abstract}
Cosmic microwave background (CMB) lensing is an integrated effect whose kernel is greater than half the peak value in the range $1 < z < 5$. Measuring this effect offers a powerful tool to probe the large-scale structure of the Universe at high redshifts. With the increasing precision of ongoing CMB surveys, other statistics than the lensing power spectrum, in particular the lensing bi-spectrum, will be measured at high statistical significance. This will provide ways to improve the constraints on cosmological models and lift degeneracies. Following on an earlier paper, we test analytical predictions of the CMB lensing bi-spectrum against full-sky lensing simulations, and discuss their validity and limitation in detail. The tree-level prediction of perturbation theory agrees with the simulation only up to $\l\sim 200$, but the one-loop order allows capturing the simulation results up to $\l\sim 600$. We also show that analytical predictions based on fitting formulas for the matter bi-spectrum agree reasonably well with simulation results, although the precision of the agreement depends on the configurations and scales considered. For instance, the agreement is at the $10\%$-level for the equilateral configuration at multipoles up to $\l\sim2000$, but the difference in the squeezed limit raises to more than a factor of two at $\l\sim2000$. This discrepancy appears to come from limitations in the fitting formula of the matter bi-spectrum. We also find that the analytical prediction for the post-Born correction to the bi-spectrum is in good agreement with the simulation. We conclude by discussing the bi-spectrum prediction in some theories of modified gravity. 
\end{abstract}

\preprint{YITP-18-127}
\maketitle


\section{Introduction} \label{sec:intro}

The path of cosmic microwave background (CMB) photons between the last scattering surface and an observer is deflected by the gravitational potential of the large-scale structure (LSS) of the Universe. This leads to distortions of the observed CMB anisotropies, which are correlated across scales. By extracting the lensing effect in these anisotropies, we can directly probe the underlying gravitational potential of the LSS, and constrain cosmological properties which affect it, such as characteristics of dark matter, dark energy, and massive neutrinos. By its nature, the lensing of CMB is sensitive to higher redshifts than most other cosmological observations of LSS, and it offers one of the most powerful probes of fundamental issues in cosmology and physics in the near future. 

Recent CMB experiments have already detected the angular power spectrum of the lensing potential very precisely \cite{ACT16:phi, BKVIII, P18:phi, PB14:phi, SPT:phi}. For instance, the {\it Planck} detection \citep{P18:phi} has a 40\,$\sigma$ significance. The detection and precise determination of the CMB lensing bi-spectrum is therefore an obvious and important next step in CMB scientific analyses. Recent studies indeed showed that the lensing potential bi-spectrum and other higher-order statistics due to non-linearity are detectable in near future CMB experiments \cite{Namikawa:2016b,Pratten:2016,Liu:2016nfs,Namikawa:2018b}. 

In previous works \cite{Namikawa:2016b,Pratten:2016}, the signal-to-noise ratio of the bi-spectrum was obtained by using the fitting formulae of Refs.~\cite{Scoccimarro:2001,Gil-Marin:2012}. However, these fitting formulae were tested in other contexts and it is unclear how they perform in this case. This paper therefore assesses the validity of these fitting formulae to calculate the CMB lensing bi-spectrum by comparing with  a full-sky CMB lensing simulation. In addition to allowing refined expectations, accurate analytic predictions for the bi-spectrum are also necessary to reduce the computational cost when analyzing real data.  

This paper is organized as follows. 
Section~\ref{sec:bispec} reviews theoretical predictions of the bi-spectrum of CMB lensing and  
Section~\ref{sec:sim} describes the simulations used in our analysis. 
Section~\ref{sec:results} presents our main results. 
Section~\ref{sec:mgresults} discusses theoretical predictions in modified gravity theories. 
Section~\ref{sec:summary} summarizes our work. 
Appendix~\ref{app:code} verifies our bi-spectrum measurements. 
Appendix~\ref{app:fit} briefly summarizes the fitting formulas of the matter bi-spectrum.  
Appendix~\ref{app:spt} explains the bi-spectrum in modified gravity theories. 
Appendix~\ref{app:squeez} further discusses the discrepancy between fitting formulae and simulations appearing in the squeezed lensing bi-spectrum. 

Throughout this paper we assume the fiducial flat $\Lambda$CDM cosmology already used in \cite{Takahashi:2017}. The cosmological parameters are $\Omega_{\rm cdm} = 0.233$, $\Omega_{\rm b} = 0.046$, $h = 0.7$, $\sigma_8 = 0.82$ and $n_{\rm s} = 0.97$. We also use natural units where $c = G = 1$.

\section{CMB lensing bi-spectrum} \label{sec:bispec}

In this section, we review briefly the general formalism of the bi-spectrum and the theoretical prediction of the bi-spectrum of CMB lensing as shown in, e.g., Refs.~\cite{Namikawa:2016b,Pratten:2016}. 

\subsection{Bi-spectrum} \label{subsec:bispec}

Denoting the harmonic coefficients of the fluctuations (e.g., the CMB temperature anisotropies) as $a_{\l m}$, the (angular) bi-spectrum is in general given by (e.g., \cite{Fergusson:2012})
\al{
	B_{\l_1\l_2\l_3} \equiv \sum_{m_1m_2m_3}\Wjm{\l_1}{\l_2}{\l_3}{m_1}{m_2}{m_3}
    	\ave{a_{\l_1m_1}a_{\l_2m_2}a_{\l_3m_3}} 
	\,. \label{Eq:bispec}
}
It is convenient to use the reduced bi-spectrum,  
\al{
	b_{\l_1\l_2\l_3} \equiv h^{-1}_{\l_1\l_2\l_3}B_{\l_1\l_2\l_3} \quad (\l_1+\l_2+\l_3 \text{ is even}) \,,
}
which removes the geometrical factor
\al{
	h_{\l_1\l_2\l_3} \equiv \sqrt{\frac{(2\l_1+1)(2\l_2+1)(2\l_3+1)}{4\pi}}
    	\Wjm{\l_1}{\l_2}{\l_3}{0}{0}{0} . 
    \label{Eq:hlll}
}
In the flat-sky approximation, the bi-spectrum is given by 
\al{
	\ave{a_{\bl_1}a_{\bl_2}a_{\bl_3}} 
		= (2\pi)^2\delta(\bl_1+\bl_2+\bl_3)\,B_{\bl_1\bl_2\bl_3}
	\,,
}
where $a_{\bl_i}$ is a Fourier mode of fluctuations. The relationship between the full- and flat-sky bi-spectrum is given by \cite{Hu:2000ee},
\al{
	B_{\l_1\l_2\l_3} &\simeq h_{\l_1\l_2\l_3} B_{\bl_1\bl_2\bl_3}
	\,.
}
This equation shows that the full-sky reduced bi-spectrum is equivalent to the flat-sky bi-spectrum. In the following, we use the flat-sky bi-spectrum to compute the reduced bi-spectrum and compare it with the results from the full sky simulation.

\subsection{Bi-spectrum from CMB lensing} \label{subsec:lensbispec}

In a CMB lensing analysis, we measure the lensing potential which is defined as (see, e.g., \cite{Lewis:2006fu,Hanson:2009kr})
\al{
	\phi(\hatn) = -2\INT{}{\chi}{}{0}{\chi_*} W(\chi,\chi_*) \Psi(\chi,\hatn) \,.
}
Here $\chi_*$ is the comoving distance to the CMB last-scattering surface and $\Psi$ is the Weyl potential. The lensing kernel, $W(\chi,\chi_*)$, is defined (for a spatially flat cosmology) as
\al{
	W(\chi,\chi_*) = \frac{\chi_*-\chi}{\chi\chi_*}\Theta(\chi_*-\chi) \,,
}
where $\Theta(x)$ denotes the Heaviside step function. The bi-spectrum of the lensing potential is then given by \eq{Eq:bispec}. Hereafter, instead of using the lensing potential, we frequently use the lensing convergence, $\kappa_{\l m}=\l(\l+1)\phi_{\l m}/2$. 

In general, a non-zero bi-spectrum is induced by non-Gaussian fluctuations, and in CMB lensing, it is sourced by two effects. One source is the non-Gaussian matter distribution induced by the nonlinear gravitational evolution \cite{Namikawa:2016b}, and the other is the next-to-leading order correction to the Born approximation for computing the deflection of the path, which is referred to as the post-Born correction \cite{Pratten:2016}. We denote these contributions respectively by $B^{\rm LSS}$ and $B^{\rm pb}$, and present their explicit expressions below.  

The bi-spectrum of the lensing convergence which arises from the nonlinear growth of the density perturbations is given in the flat-sky limit by \cite{Namikawa:2016b}
\al{
	B^{\rm LSS}_{\bl_1\bl_2\bl_3} = \INT{}{\chi}{}{0}{\chi_*} &
		\left[\frac{3\Omega_{m,0}H_0^2}{2a(\chi)}\right]^3
		\notag \\
		\times &\ \chi^2 W^3(\chi,\chi_*)\, B_\delta(\bk_1,\bk_2,\bk_3,\chi)
	\,, \label{Eq:bisp:lss}
}
where $\bk_i=\bl_i/\chi$. Here, $B_\delta$ is the
\emph{matter} bi-spectrum which results from the nonlinear growth of structure. In the weakly nonlinear regime, it can be obtained by using perturbation theory. The result at the tree-level order is of the general form
\al{
	B_\delta(\bk_1,\bk_2,\bk_3,\chi) 
		&= 2F_2(\bk_1,\bk_2,z)\,P_{\delta}(k_1,z)P_{\delta}(k_2,z) 
	\notag \\
		&+ 2\,{\rm perms.}
	\, ,
	\label{eq:B_LSS_tree}
}
where $P_\delta(k,z)$ is the matter power spectrum at redshift $z(\chi)$, and the function $F_2$ is the second-order perturbation theory kernel (e.g., \cite{Bernardeau:2001qr}). Writing $\bk_1\cdot\bk_2=k_1k_2\cos\theta$, it is given by
\al{
	F_2(\bk_1,\bk_2,z) &= \frac{5}{7}\,a(k_1,z)a(k_2,z) 
		\notag \\
		&+ \frac{1}{2}\,\frac{k_1^2+k_2^2}{k_1k_2}\,b(k_1,z)b(k_2,z)\cos\theta 
		\notag \\
		&+ \frac{2}{7}\,c(k_1,z)c(k_2,z)\cos^2\theta 
	\,, \label{Eq:F2:GR}
}
where $a(k,z)$, $b(k,z)$ and $c(k,z)$ are unity at the tree-level in perturbation theory. In order to capture the deviation from the tree-level prediction, one may calculate the correction at next-to-leading order (one-loop) to the prediction. We will briefly discuss its validity in subsequent analysis (see Sec.~\ref{subsec:bk_1loop}). In this paper, our primary focus is to compare the lensing simulations with analytical predictions based on fitting formulae of the matter bi-spectrum. These fits provide an analytical functional form for $a(k,z)$, $b(k,z)$ and $c(k,z)$. We will assess the most recent results for these fitting formulae, as given by Gil-Marin et al. in  Ref.~\cite{Gil-Marin:2012}, hereafter denoted by ``GM", and the earlier results of Scoccimaro and Couchman \cite{Scoccimarro:2001}, denoted below by ``SC". 

Furthermore, the post-Born correction to the CMB lensing bi-spectrum, $B^{\rm pb}$ is expressed, as in \cite{Pratten:2016}, by
\al{
	B^{\rm pb}_{\bl_1\bl_2\bl_3} 
		&= 2\frac{\bl_1\cdot\bl_2} {\l^2_1\l^2_2}\left[\bl_1\cdot\bl_3\, M_{\l_1\l_2}+\bl_2\cdot\bl_3\,M_{\l_2\l_1}\right] 
	\notag \\
		&+ {\rm cyc. perm.} 
	\,, \label{Eq:bisp:pb}
}
where 
\al{
	M_{\l\l'} &= (\l\l')^4\INT{}{\chi}{}{0}{\chi_*} \frac{[W(\chi,\chi_*)]^2}{\chi^2}P_\Psi \left(\frac{\l}{\chi},\chi\right) 
		\notag \\
		&\times \INT{}{\chi'}{}{0}{\chi}\frac{W(\chi',\chi)W(\chi',\chi_*)}{(\chi')^2}
		P_\Psi\left(\frac{\l'}{\chi'},\chi'\right)
	\,, 
}
$P_\Psi(k,\chi)$ being the power spectrum of the Weyl potential at a comoving distance $\chi$. 

\begin{figure*}[htb]
\bc
\includegraphics[width=89mm,clip]{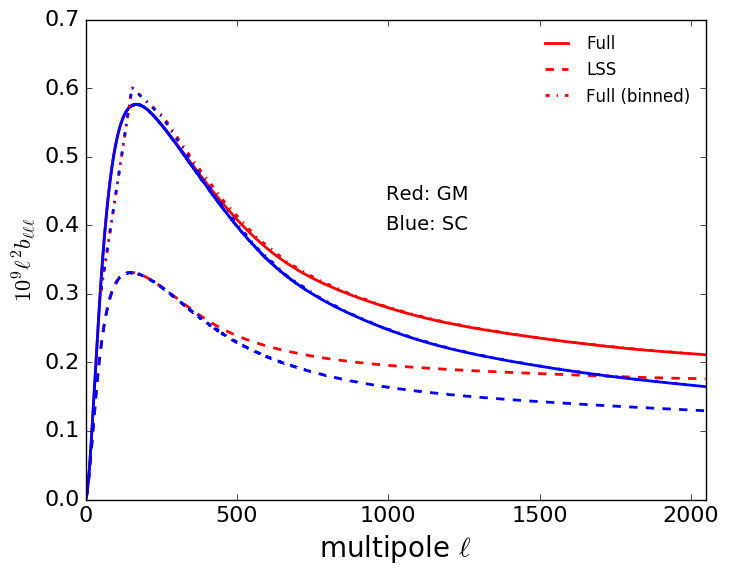}
\includegraphics[width=89mm,clip]{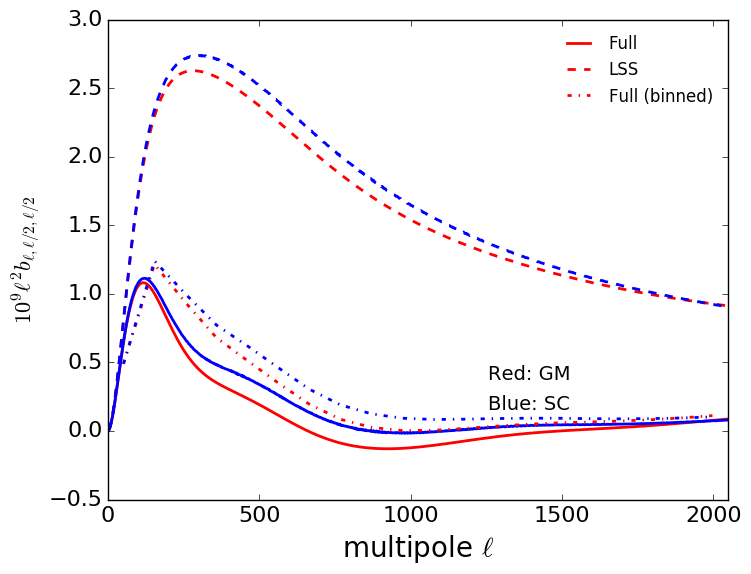}
\includegraphics[width=89mm,clip]{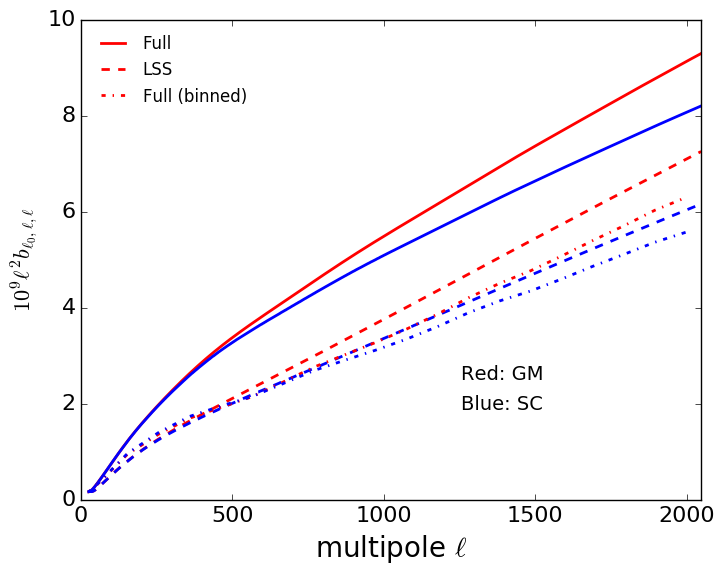}
\includegraphics[width=89mm,clip]{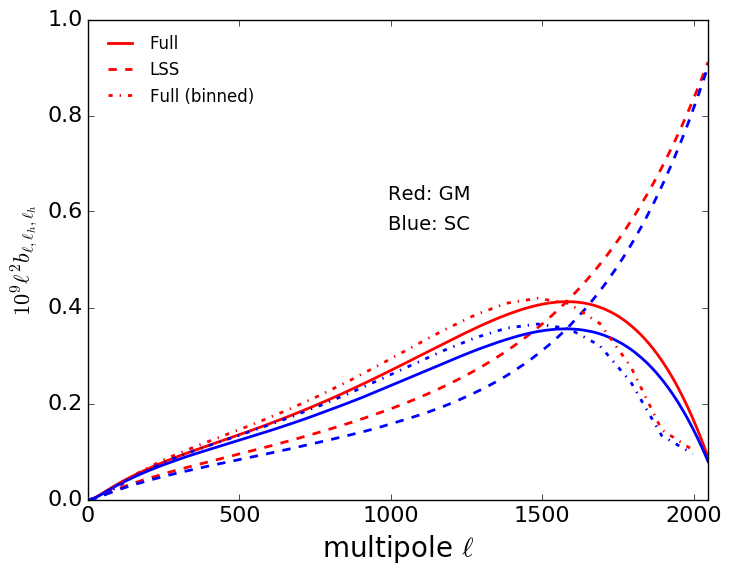}
\caption{
Reduced bi-spectra, as computed from \eqs{Eq:bisp:lss,Eq:bisp:pb}, in various configurations. Equilateral ({ Top Left}), folded ({ Top Right}), squeezed ({ Bottom Left}) and isosceles ({ Bottom Right}). The dashed lines show the contribution from  the matter bi-spectrum alone \eq{Eq:bisp:lss}. The dot-dashed lines show the binned bi-spectrum defined in \eq{Eq:bispec:binned:def} with $20$ multipole bins between $\l=1$ and $2048$. 
}
\label{fig:bispec:predict}
\ec
\end{figure*}

Fig.~\ref{fig:bispec:predict} shows the theoretical predictions of the CMB lensing bi-spectrum given by \eqs{Eq:bisp:lss,Eq:bisp:pb}, assuming either the GM or SC fitting formulae, for the following four specific configurations: \begin{enumerate}
    \item[C1.] equilateral, with $\l_1=\l_2=\l_3=\l$, 
    \item[C2.] folded, with $\l_1=2\l_2=2\l_3=\l$, 
    \item[C3.] squeezed, with $\l_1=50$, $\l_2=\l_3=\l$, and
    \item[C4.] isosceles, with $\l_1=\l$, $\l_2=\l_3=1000$.
\end{enumerate}
We also show the contribution from the matter bi-spectrum alone. This figure clearly shows that the post-Born correction is important and changes the shape of the bi-spectrum in all of the above configurations. 

The figure also shows the effect of multipole binning, which is discussed further in Sec.~\ref{subsec:measurement}. Binning is not necessary to constrain cosmological parameters, but in practice, it is rather essential, e.g., to reduce the computational cost. Here, we adopt binning in order to reduce the dispersion of the simulation measurements, as follows: 
\al{
	b_{b_1b_2b_3} \equiv \frac{\sum_{\l_i}^{b_i}h^2_{\l_1\l_2\l_3}b_{\l_1\l_2\l_3}}{\sum_{\l_i}^{b_i}h^2_{\l_1\l_2\l_3}} 
    \,, \label{Eq:bispec:binned:def}
}
where $\sum_{\l_i}^{b_i}$ ($i=1,2,3$) means the summation over $\l_i$ within the multipole bin, $b_i$. As shown in Fig.~\ref{fig:bispec:predict}, the binned bi-spectrum (dot-dashed) can be very different from the unbinned bi-spectrum (solid). For example, the equilateral binned bi-spectrum, $b\equiv b_1=b_2=b_3$, contains not only the unbinned equilateral bi-spectrum but also all of the other bi-spectra which satisfies $\l_1,\l_2,\l_3\in b$. The figures show that the binning effect is not significant in the equilateral case but crucial in other configurations.

\section{CMB lensing bi-spectrum in simulations} \label{sec:sim}

\subsection{Full-sky ray-tracing simulations}

In this subsection, we briefly describe the full-sky ray-tracing simulations which we used to construct the lensing maps. Details may be found in Ref.~\cite{Takahashi:2017}. First, we performed cosmological $N$-body simulations of dark matter to produce an inhomogeneous mass distribution in the Universe, which properly describe the non-Gaussian character of the matter distribution. The simulations followed the gravitational evolution of $2048^3$ particles in cubic simulation boxes by using the public code Gadget2~\cite{Springel:2001,Springel:2005}. The cubic boxes have side lengths of $450, \, 900, \, 1350,\dots, 6300 \, h^{-1} {\rm Mpc}$ from low to high redshift $(z \leq 7.1)$. We chose an arbitrary point in the box as the observer's position and constructed spherical lens shells with a thickness of $150h^{-1} \, {\rm Mpc}$ around the observer (who is located at the center of the shells). 
We construct three lens shells (having a $450h^{-1} \, {\rm Mpc}$ total thickness) in each box. The inner shell is taken from the smaller box at lower redshift.
Then, we projected the $N$-body particles onto each lens shell, and calculated the surface mass density and deflection angle. For higher redshifts ($z=7.1-1100$), we prepared the lens shells based on the Gaussian fluctuations (instead of the $N$-body method), which is a good approximation in the linear regime. Light rays, emitted from the observer, are deflected at each lens shell, and these ray paths are numerically evaluated up to the last scattering surface. Therefore, the simulation includes the post-Born correction. We used the public code {\tt GRayTrix}\footnote{\url{http://th.nao.ac.jp/MEMBER/hamanatk/GRayTrix/}} \cite{Hamana:2015,Shirasaki:2015} for this ray-tracing computation. The code relies on the {\tt Healpix} scheme \cite{Gorski:2004by}. We adopt the angular resolution of $N_{\rm side}=4096$ in our main results (but we also increase the resolution to $N_{\rm side}=8192$ in order to see the dependence of our results on the map resolution). We numerically obtain the CMB deflection angle $\bm{d}(\hatn)$ on the full sky maps, and then compute the harmonic coefficients of the lensing potential, $\grad_{\l m}$, using the spin-1 harmonic transform. The harmonic coefficients of the lensing convergence are finally obtained by $\kappa_{\l m}=\l(\l+1)\grad_{\l m}/2$. We prepared $108$ such maps in total. We checked that the angular power spectrum of the lensing potential agrees with the theoretical prediction calculated by {\tt CAMB} \cite{Lewis:1999bs} within $5\%$ up to $\l=3000$ (see Section 3.5 of Ref.~\cite{Takahashi:2017}). 

In addition to the full-sky CMB lensing maps which include both the nonlinear growth and the post-Born effect, we also created CMB lensing maps only including the latter. Such maps are useful to study each contribution separately. To create the maps, we construct lens shells based on the Gaussian fluctuations without modifying the density power spectrum. Then, we performed the ray-tracing simulation, and repeated the same simulation with a different realization to obtain 10 maps.

\subsection{Measurement of the lensing bi-spectrum}
\label{subsec:measurement}

Next we describe how we measure the binned lensing reduced bi-spectrum in practice. The full-sky reduced (unbinned) bi-spectrum, $b_{\ell_1\ell_2\ell_3}$, is rewritten with 
\begin{widetext}
\al{
	b_{\l_1\l_2\l_3} &= h^{-1}_{\l_1\l_2\l_3} \sum_{m_1m_2m_3} 
		\Wjm{\l_1}{\l_2}{\l_3}{m_1}{m_2}{m_3} \ave{\kappa_{\l_1m_1}\kappa_{\l_2m_2}\kappa_{\l_3m_3}} 
        \notag \\
 	   &= h^{-1}_{\l_1\l_2\l_3} \sum_{m_1m_2m_3}
       	h^{-1}_{\l_1\l_2\l_3}\Int{2}{\hatn}{} Y_{\l_1m_1}(\hatn)Y_{\l_2m_2}(\hatn)Y_{\l_3m_3}(\hatn)
        \ave{\kappa_{\l_1m_1}\kappa_{\l_2m_2}\kappa_{\l_3m_3}}
        \notag \\
 	   &= h^{-2}_{\l_1\l_2\l_3} \AVE{\Int{2}{\hatn}{} \kappa_1(\hatn)\kappa_2(\hatn)\kappa_3(\hatn)}
    \,, \label{Eq:bispec:sim}
}
\end{widetext}
where $\kappa_i(\hatn)$ is obtained by the inverse harmonic transform of $\delta_{\l\l_i}\kappa_{\l m}$. From the first to the second line, we use the formula of the Gaunt integral shown in \eq{Eq:gaunt}. The square of the geometrical factor is rewritten as 
\begin{widetext}
\al{
	h^2_{\l_1\l_2\l_3} &= \frac{(2\l_1+1)(2\l_2+1)(2\l_3+1)}{4\pi}\Wjm{\l_1}{\l_2}{\l_3}{0}{0}{0}^2
    	\notag \\
    &= \sqrt{\frac{(2\l_1+1)(2\l_2+1)(2\l_3+1)}{4\pi}}
    	\Int{2}{\hatn}{} Y_{\l_10}(\hatn)Y_{\l_20}(\hatn)Y_{\l_30}(\hatn)
    	\notag \\
    &= \frac{1}{\sqrt{4\pi}} \Int{2}{\hatn}{} h_1(\hatn)h_2(\hatn)h_3(\hatn)
    \,, \label{Eq:bispec:norm}
}
\end{widetext}
where $h_i(\hatn)$ is given by the inverse harmonic transform of $\sqrt{2\l_i+1}\delta_{\l\l_i}\delta_{m0}$. 

To reduce the simulation error from the cosmic variance, we measure the binned reduced bi-spectrum of \eq{Eq:bispec:binned:def} from the simulation data. Substituting \eqs{Eq:bispec:sim,Eq:bispec:norm} into \eq{Eq:bispec:binned:def}, we find 
\al{
	b_{b_1b_2b_3} &\simeq \sqrt{4\pi}\frac{\sum_p \kappa_1(\hatn_p)\kappa_2(\hatn_p)\kappa_3(\hatn_p)}
    {\sum_p h_1(\hatn_p)h_2(\hatn_p)h_3(\hatn_p)}
    \,, \label{Eq:bispec:binned}
}
where $p$ is the pixel index, and $\sum_p$ is the sum over all pixels. The discretization of the sphere applied to the above equation is tested in Appendix \ref{app:code}.

\section{Results} \label{sec:results}

\begin{figure*}[htb]
\bc
\includegraphics[width=89mm,clip]{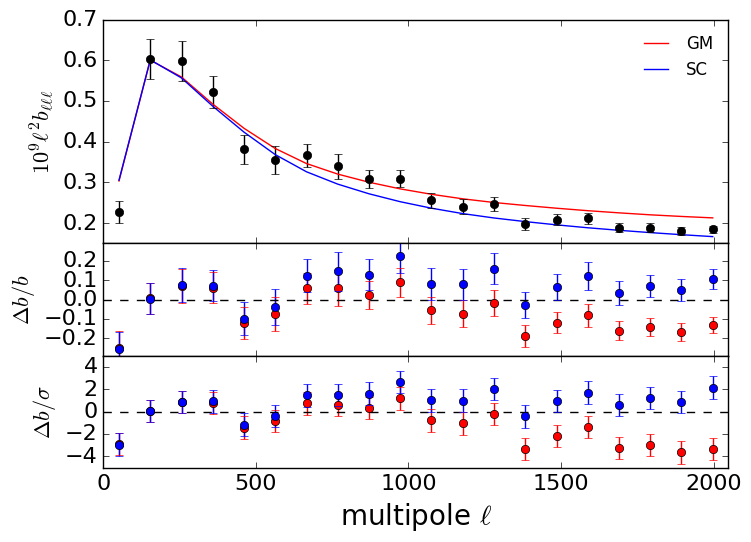}
\includegraphics[width=89mm,clip]{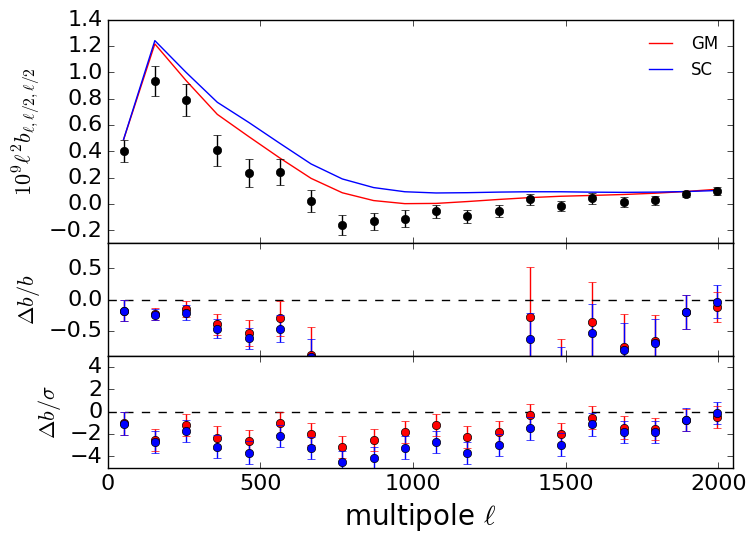}
\includegraphics[width=89mm,clip]{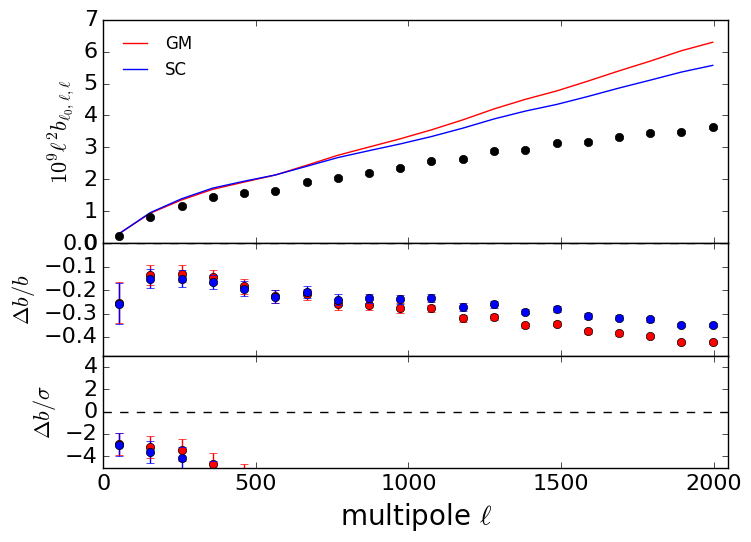}
\includegraphics[width=89mm,clip]{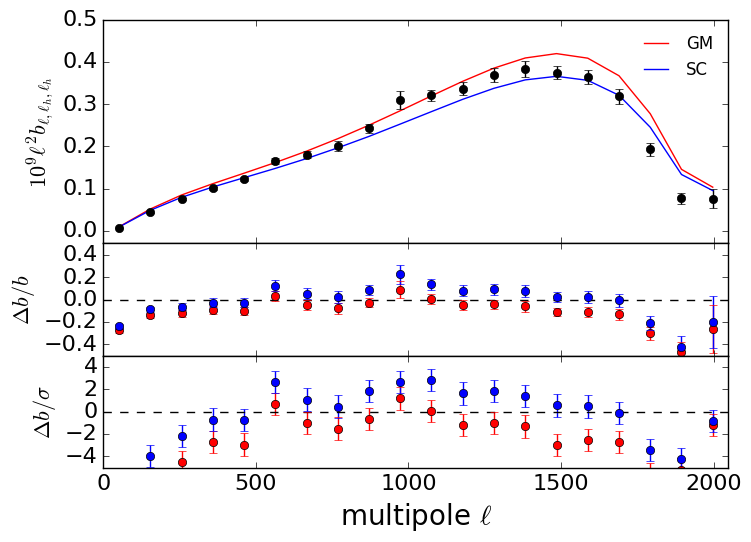}
\caption{
Equilateral ({\bf Top Left}), folded ({\bf Top Right}), squeezed ({\bf Bottom Left}) and isosceles ({\bf Bottom Right}) configurations of the reduced bi-spectra measured from simulation (solid points) compared with theoretical models using either the SC (blue) and GM (red) fitting formulae. The top panels show the bi-spectra with the simulation points in black, the middle panels show the fractional differences, $\Delta b/b$  of the simulations points, coloured according to the reference model they are compared to, and the bottom panels rather show difference normalised by the simulation uncerrtainties, $\Delta b/\sigma$. 
}
\label{fig:bispec:b20}
\ec
\end{figure*}

In this section, we compare analytic predictions of the lensing bi-spectrum with measured results from lensing  simulations, focusing mainly on the predictions based on fitting formula. We first present the results of the total lensing bi-spectrum in Sec.~\ref{subsec:bk_total}. Then, the contribution from post-Born correction is measured separately and compared with analytic predictions in Sec.~\ref{subsec:bk_post-Born}. Based on this, possible reasons for the discrepancy seen in Sec.~\ref{subsec:bk_total} are discussed. In Sec.~\ref{subsec:bk_1loop}, we examine the perturbation theory calculation, and the applicable range of higher-order (one-loop) predictions is discussed.

\subsection{Total contributions} 
\label{subsec:bk_total}

Fig.~\ref{fig:bispec:b20} compares theory and simulations for the same four configurations presented in Fig~\ref{fig:bispec:predict} (equilateral, folded, squeezed and isosceles). The analytic prediction is computed based on \eqs{Eq:bisp:lss,Eq:bisp:pb}. Note that, in the simulation, we compute the bi-spectrum of the lensing convergence instead of the lensing potential. The top panel in each figure shows the bi-spectrum, the middle panel shows the fractional difference, $\Delta b/b\equiv (b^{\rm sim}-b^{\rm theory})/b^{\rm theory}$, and the bottom panel shows the difference against the simulation dispersion, $\Delta b/\sigma\equiv (b^{\rm sim}-b^{\rm theory})/\sigma$. 

In the equilateral case, the analytic prediction with the SC fitting formula agrees with the simulation result, at least, up to $\l=2048$ within the simulation error. The analytic result with the GM fitting formula slightly overestimates the simulation results. In the folded case, the analytic results with both the SC and GM fitting formulas overestimate the simulation at almost all scales. The most significant discrepancy is found in the squeezed configuration. The analytic predictions start to overestimate the simulation results as we go to small scale and the discrepancy is much larger than the simulation error. The fractional difference is roughly $30-40$\% at small scales ($\l\geq 1000$). In the isosceles case, the analytic prediction becomes consistent with the simulation if $\l_1$ becomes close to $\l_2=\l_3=1000$. 

\begin{figure*}[t]
\bc
\includegraphics[width=89mm,clip]{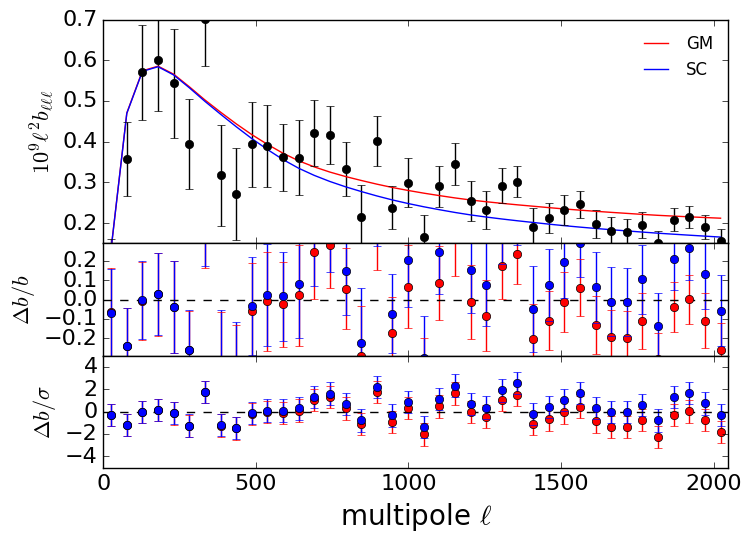}
\includegraphics[width=89mm,clip]{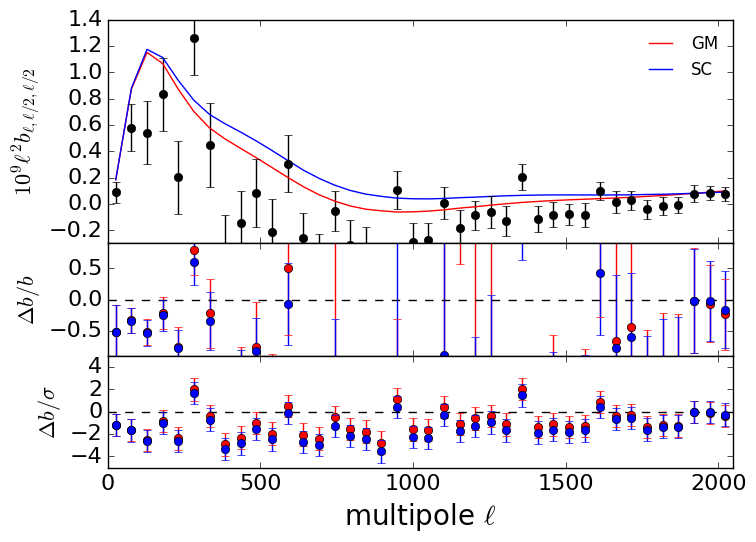}
\includegraphics[width=89mm,clip]{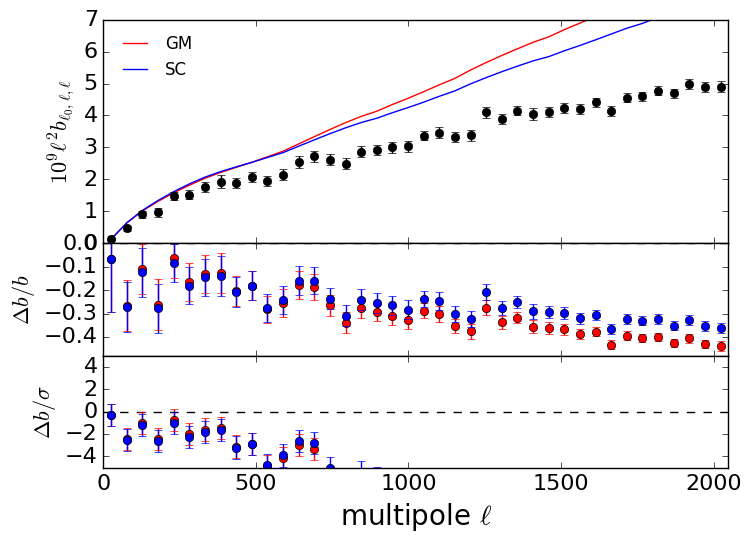}
\includegraphics[width=89mm,clip]{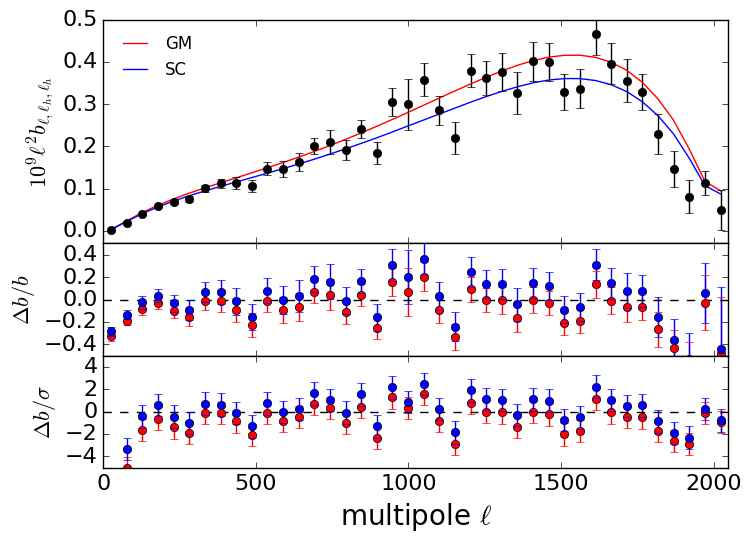}
\caption{
Same as Fig.~\ref{fig:bispec:b20} but employing $40$ multipole bins instead of 20. 
}
\label{fig:bispec:b40}
\ec
\end{figure*}

In order to see the dependence on the size of the multipole bins, Fig.~\ref{fig:bispec:b40} shows the bi-spectrum with an increased number of multipole bins. If the number of multipole bins increases, the binned bi-spectrum is less contaminated by different configurations of the bi-spectrum while the simulation error increases. The consistency between the analytic and simulated bi-spectrum is improved compared to the case with $20$ multipole bins, mostly due to the increase of the error bars. However, a huge discrepancy in the squeezed configuration still remains at small scales, and its size is almost the same as that shown in Fig.~\ref{fig:bispec:b20} (see also Fig.~\ref{fig:nres} in Appendix \ref{app:squeez}). 

\begin{figure*}[t]
\bc
\includegraphics[width=89mm,clip]{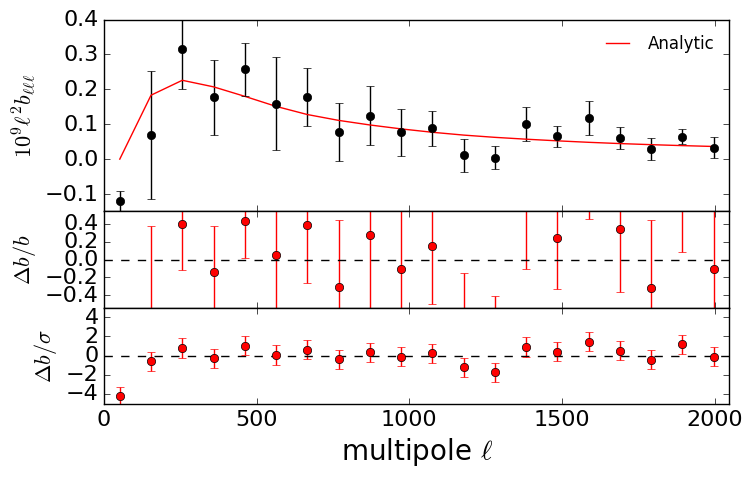}
\includegraphics[width=89mm,clip]{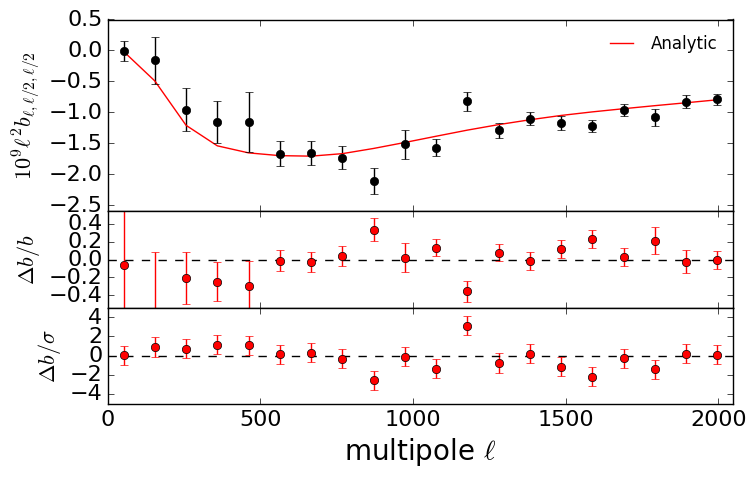}
\includegraphics[width=89mm,clip]{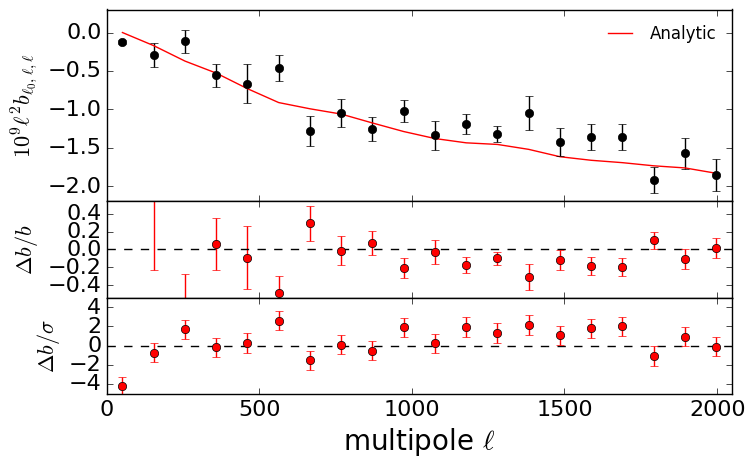}
\includegraphics[width=89mm,clip]{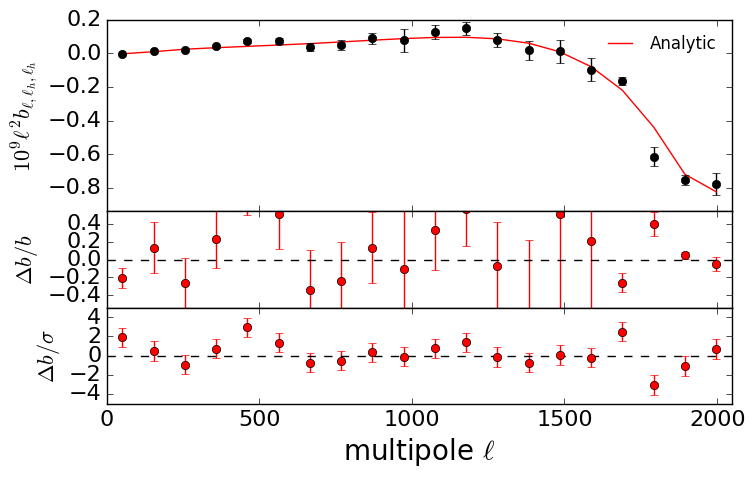}
\caption{
Same as Fig.~\ref{fig:bispec:b20} but for the contributions only from the post-Born effect. We average over the 10 realizations of the post-Born simulation. 
}
\label{fig:bispec:postborn}
\ec
\end{figure*}

\subsection{Post-Born correction}
\label{subsec:bk_post-Born}

As shown in Refs.~\cite{Pratten:2016,Lewis:Pratten:2016}, the post-Born correction changes the shape of the bi-spectrum. Nevertheless, their discussions are based on the leading-order contributions to the bi-spectrum, and the validity and accuracy of their predictions have not yet been tested. Here, we compare the leading-order analytic prediction with the post-Born simulation. Fig.~\ref{fig:bispec:postborn} shows the bi-spectrum measurements from the post-Born simulation to check the consistency between the analytic predictions and simulation results. The analytic prediction of the post-Born bi-spectrum is given by \eq{Eq:bisp:pb}, with $P(k)$ computed using the HALOFIT prescription~\cite{Smith:2002dz,Takahashi:2012em}. The leading-order predictions are indeed in good agreement with the simulation results, and no notable discrepancy is found over all scales, suggesting that the leading-order contribution is sufficient to describe post-Born correction to the lensing bi-spectrum.  

As we discuss in Sec.~\ref{subsec:bk_total}, although the prediction by the analytic fitting formula is consistent with the simulation results in most configurations and scales, we find a significant discrepancy in the squeezed configuration (a similar result was found in Ref.~\cite{Coulton:2018ebd} using a flat-sky simulation). 
{The good match of the post-Born prediction with simulation measurements seen in Fig.~\ref{fig:bispec:postborn} suggests that the discrepancy mostly comes from the LSS contribution, as theoretically predicted by \eq{Eq:bisp:lss}. Further, the close match of the post-Born prediction and simulation measurements also suggests the discrepancy observed is not a failing of the Limber approximation. Indeed, we find a discrepancy between the squeezed matter bi-spectra obtained from simulation and fitting formula especially at low $z$, i.e., the fitting formula is not accurate in the squeezed limit. This may be because of the configurations and scales the formula was fit to. This may also be partly because the fitting formula is derived using the unbinned calculation while the binning effect significantly changes the squeezed bi-spectrum prediction. Ref.~\cite{Coulton:2018ebd} found that the poor resolution of the simulation map leads to a discrepancy in the small-scale squeezed bi-spectrum, but our results are not significantly affected by the map resolution. While a thorough exploration of the discrepancy requires the validation of the fitting formula at all redshifts and scales and is beyond the scope of this paper, we further discuss possible sources of the discrepancy in Appendix \ref{app:squeez}. 
}

\begin{figure*}[t]
\bc
\includegraphics[width=89mm,clip]{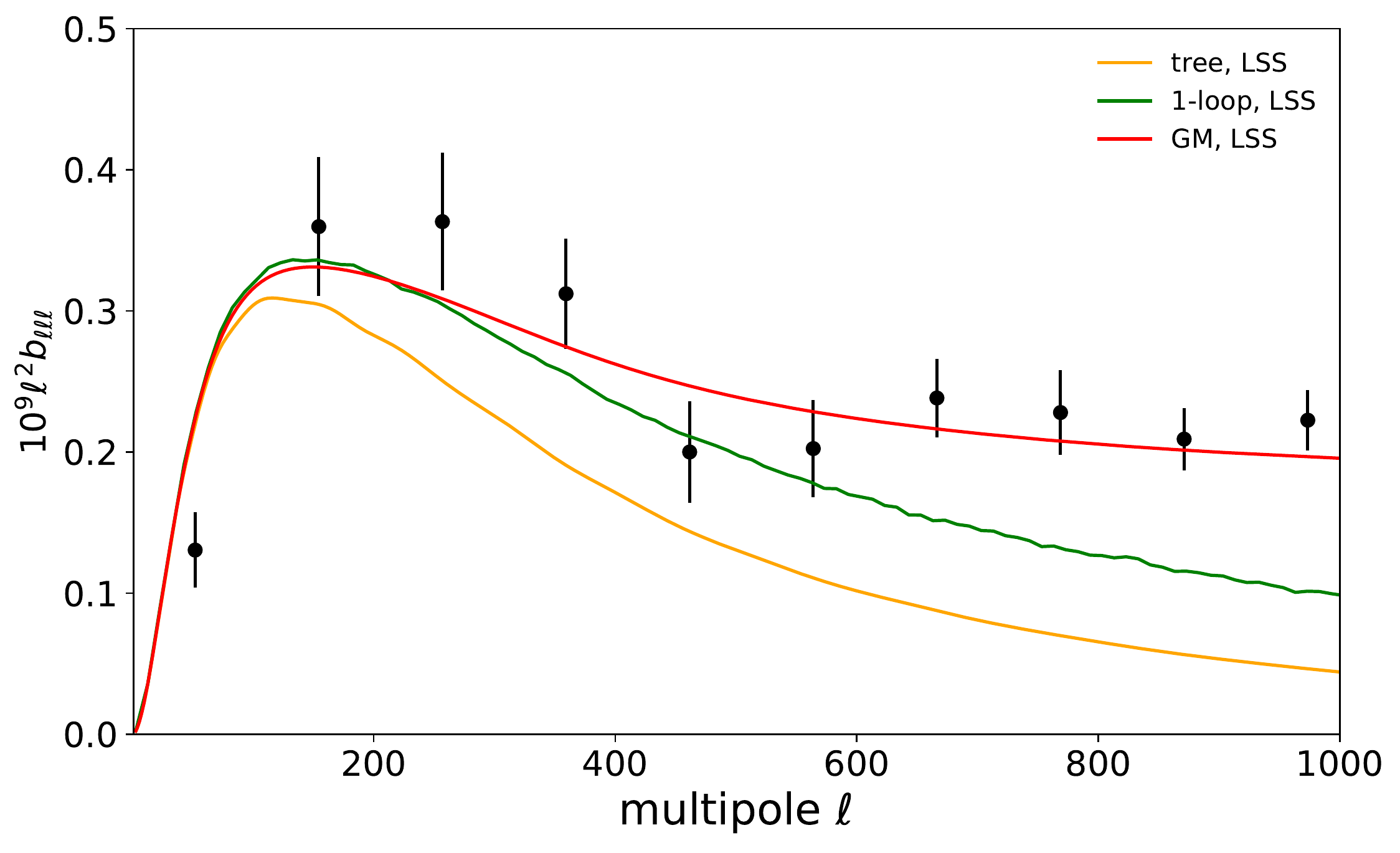}
\includegraphics[width=89mm,clip]{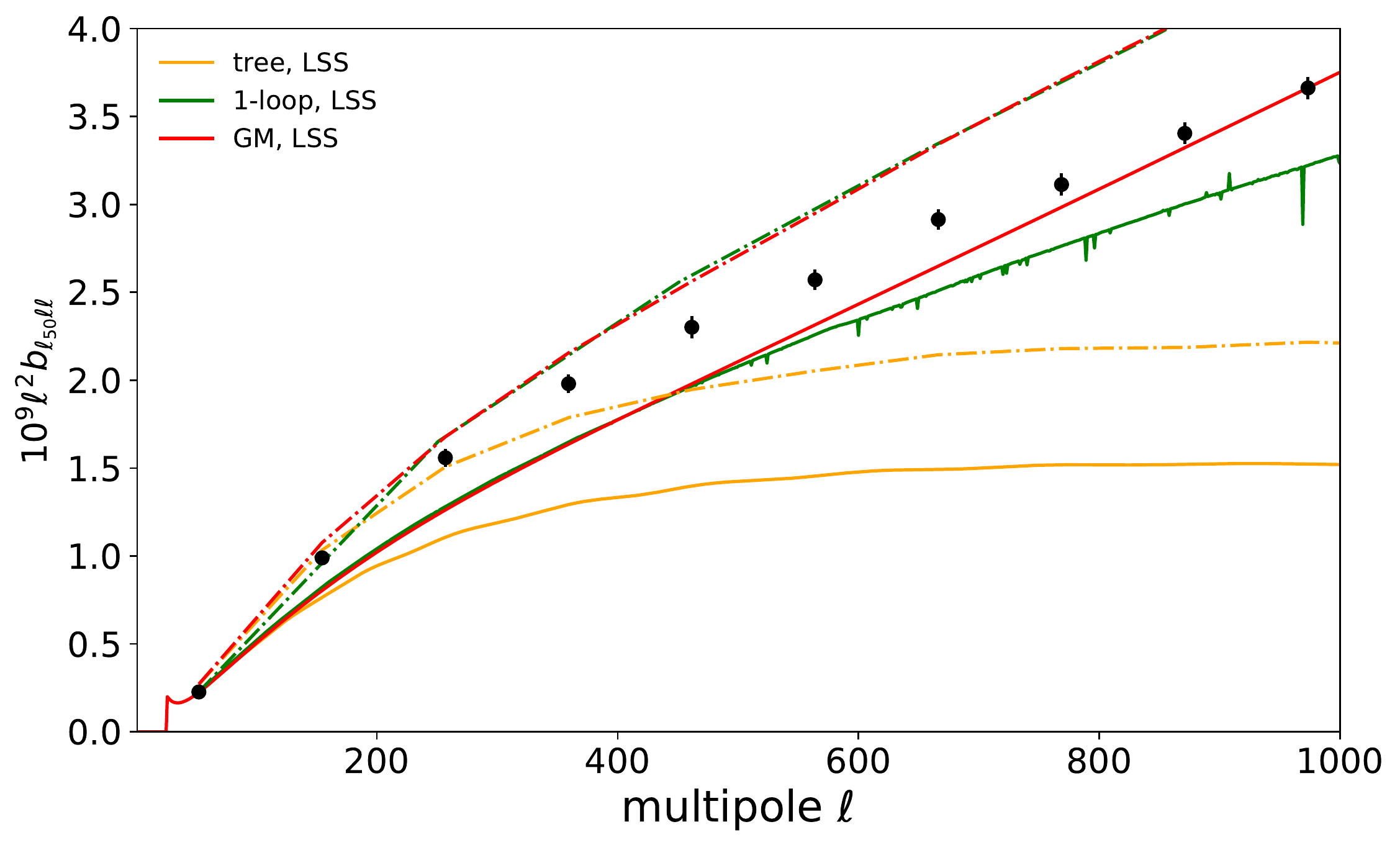}
\caption{
Equilateral (left) and squeezed (right) LSS contributions to CMB lensing bi-spectrum measured from simulation (black points) compared with tree (orange), one-loop (green) and Gil-Marin fitting formula (red) predictions. The solid lines depict the unbinned result while the dot-dashed lines depict the binned result using 20 linearly spaced bins in the range $1 \leq \ell \leq 2048$. 
}
\label{fig:regptvhalo}
\ec
\end{figure*}

\subsection{Comparison with perturbation theory}
\label{subsec:bk_1loop}

The analytic predictions of the lensing bi-spectrum have up to this point been computed with fitting formula for the matter bi-spectrum. Here, we examine the range of validity of perturbation theory calculations, specifically we compare the prediction based on one-loop order calculations with our simulation measurements. Fig.~\ref{fig:regptvhalo} shows the results for the equilateral and squeezed configurations. Note that only the LSS contribution as given by \eq{Eq:bisp:lss} is shown, with the black points being computed by subtracting the post-Born correction from the simulation measurement. The analytic predictions for the equilateral configuration (left panel) are computed without the binning effect since the impact of the multipole binning is negligible in this case as shown in Fig.~\ref{fig:bispec:predict}. We find the one-loop prediction agrees with the simulation up to $\l \sim 600$ and does substantially better than linear theory although valuable non-linear information is still missing from this prediction. In the right panel, the predictions for the squeezed limit configuration are plotted, with the binned predictions depicted as dot-dashed lines. We find that all predictions fail to trace the simulation measurements for $300 \leq \ell$ (see Appendix \ref{app:squeez} for more details). While the binned predictions from  the fitting formula and one-loop calculation are indistinguishable in the binned case at $\ell \leq 1000$, the  unbinned results show a significant discrepancy at higher $\ell$ when unbinned. 

Albeit its limited applicability range, one advantage of the perturbation theory approach is its  flexibility. Within the framework of  perturbative calculation, one can construct predictions for general theories of gravity and dark energy with relatively few assumptions thrown into the modelling. In the next section, based on the range of applicability deduced here, we use one-loop perturbation theory to investigate the signal of modified gravity on the CMB lensing bi-spectrum.

\vspace*{0.5cm}

\section{Signal of modified gravity} \label{sec:mgresults}

\begin{figure*}[t]
\bc
\includegraphics[width=89mm,clip]{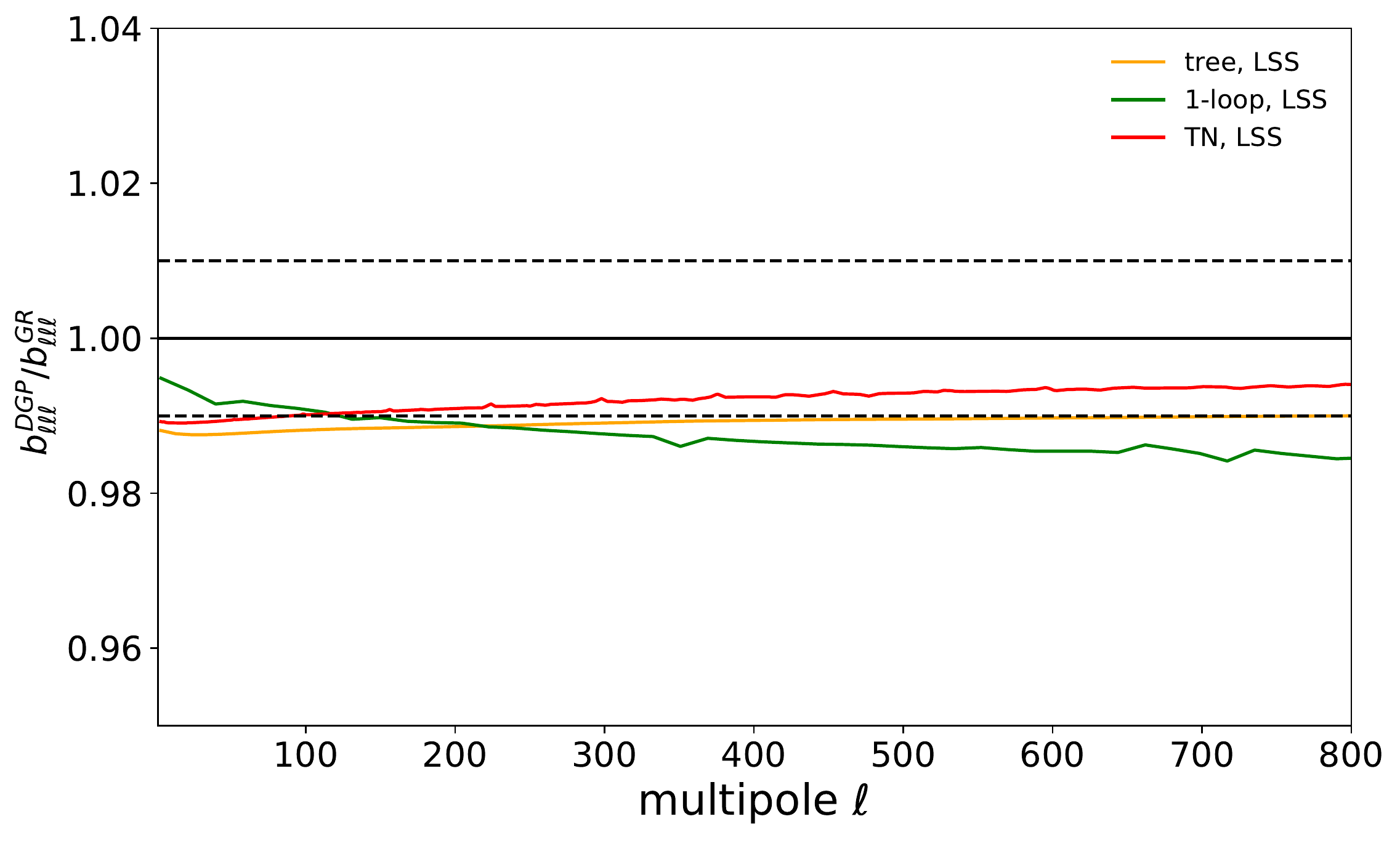}
\includegraphics[width=89mm,clip]{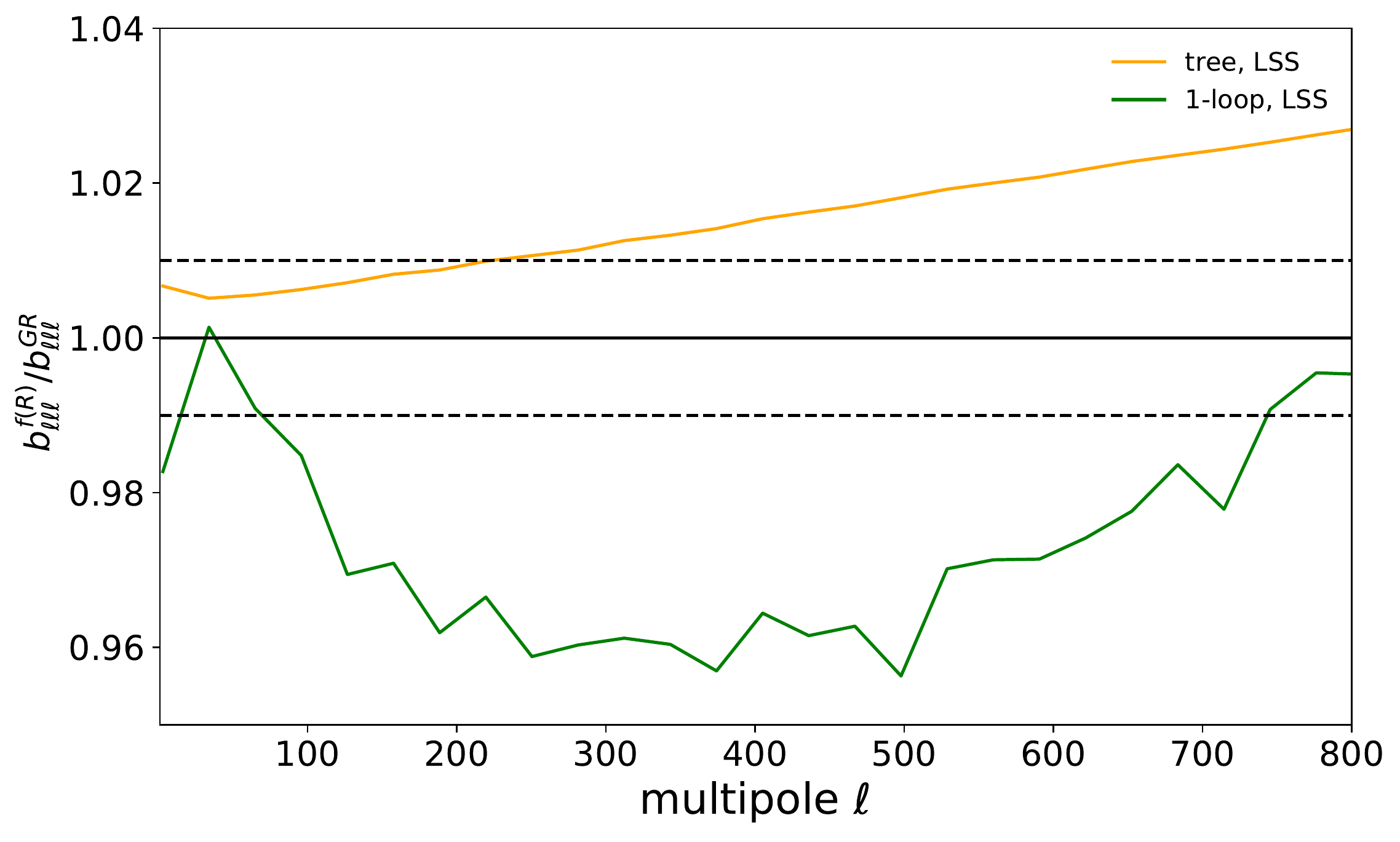}
\caption{
Ratio of equilateral LSS contribution to the CMB lensing bi-spectrum in DGP model (left) and $f(R)$ gravity (right) to the GR prediction for various theoretical predictions. For DGP we assume $\Omega_{\rm rc}=0.438$, and for $f(R)$ we assume $|\bar{f}_{R0}| = 2.5 \times 10^{-6}$. 
}
\label{fig:mg}
\ec
\end{figure*}

In this section, we investigate the impact of modified gravity on the CMB lensing bi-spectrum. Beyond GR, currently available predictions of lensing bispectrum, especially for LSS contribution, are from the perturbation theory calculation through the prediction of matter bi-spectrum that has been recently extended by Ref.~\cite{Bose:2018zpk}, based on the general framework of perturbation theory treatment (\cite{Koyama_etal2009,Taruya2016,Bose:2016qun}). No fitting formula for matter bi-spectrum has been calibrated from modified gravity simulations. Nevertheless, a simple treatment proposed by Ref.~\cite{Namikawa:2018b} is to stick to the the fitting formula of matter bi-spectrum in Eq.~(\ref{eq:B_LSS_tree}) with the kernel function given by Eq.~(\ref{Eq:F2:GR}), but introducing some freedoms in the kernel function. With this modification, the fitting formula allows us to predict the lensing bi-spectrum in a rather wide class of modified  theories of gravity, including Horndeski theory\cite{Horndeski1974} and beyond-Horndeski class theory \cite{Gleyzes:2015a,Gleyzes:2015b,Langlois:2016a,Langlois:2016b}. Indeed, Ref.~\cite{Bose:2018zpk} has checked with modified gravity simulations that this treatment provide a reasonable description of matter bi-spectrum at large scales in several modified gravity models.

Here, we restrict our analysis to two representative models of modified gravity, both with distinct screening mechanisms which are essential theoretical implementations that recover GR at small scales. This allows them to be consistent with solar system tests of gravity. The representative models we choose are the Vainshtein-screened DGP model of gravity \cite{Dvali:2000hr} and the chameleon-screened Hu-Sawicki $f(R)$ model of gravity \cite{Hu:2007nk}. The DGP model also serves as a representative of the Horndeski class of models and so we may look at two distinct predictions, that given by perturbation theory and that given by the fitting formula described in Ref.~\cite{Namikawa:2018b}. On the other hand, the $f(R)$ case has no non-linear matter bi-spectrum fitting formula available in the literature, and so only a perturbative treatment will be considered. For the perturbative treatment we consider both the tree level and one-loop prediction (see Appendix \ref{app:spt}). The theory parameters we choose are $\Omega_{\rm rc}=0.438$\footnote{A common parametrization of the cross-over scale, see \cite{Bose:2016qun} for example.} and $|\bar{f}_{R0}|=2.5\times10^{-6}$ which is a recent constraint coming from the LSS observations \cite{Smith:2009fn}. 

In Fig.~\ref{fig:mg}, we show the ratio of the LSS contribution (to the CMB lensing bi-spectrum) in DGP (left) and $f(R)$ gravity (right) to the prediction of GR. We consider the equilateral configuration as it has been shown to offer the largest signal of modified gravity in the representative models chosen \cite{Bose:2018zpk} as well as the theoretical predictions giving good consistency with simulation measurements (see Fig.~\ref{fig:bispec:b20}). In both models, the linear power spectrum is identical to GR at large scales.\footnote{Note that in the DGP case we also have a modification to the linear growth factor which we normalize to that of GR. This factor is completely degenerate with $\sigma_8$ and so such scale independent modifications to the linear power spectrum are not very useful to test modified gravity unless one can break such a degeneracy using other probes.}
For the DGP case, the deviation shown here comes only from non-linear screening effects while for $f(R)$ gravity we have effects both from screening and scale-dependent modifications to the linear growth of structure. In both panels, we show the tree-level prediction (orange) and one-loop prediction (green).\footnote{The noise seen in the plots is numerical.}
For the DGP case we also show the prediction coming from the non-linear fitting formula of Ref.~\cite{Namikawa:2018b} (red). 

Note that the effect of screening mechanism is perturbatively incorporated into the perturbation theory predictions through the modified higher-order kernels. Because of this, the screening effect is ineffective at tree-level order, and the predictions do not approach unity (i.e., GR) at small scales. On the other hand, the one-loop predictions depicted as green curves include the screening effect up to the fourth-order kernels in perturbation theory. These higher-order corrections give distinct scale-dependent shapes and dominate the signal at $\l>100$. They also work to reduce the linear modification. This is not evident in the DGP case (left panel) as we have normalized the linear growth of DGP to that of GR. Rather, this is clearer in the $f(R)$ case (right). Additional orders in perturbation theory would work to complete the screening at small scales. On this point, the fitting formula shown for the DGP case deviates from the one-loop calculation noticeably at smaller scales. This can be expected as the effect of screening mechanism is not fully incorporated into the fitting formula.

Fig.~\ref{fig:mg} indicates that the impact of the modification to gravity we considered is of the order of a few percent at most. The CMB Stage-IV experiments would not be able to probe such small signals of modified gravity at the multipoles considered \cite{Namikawa:2018b}. However, this is only the case if we restrict the scales to the applicable range of one-loop perturbation theory. At smaller scales, a rather notable deviation from GR is still allowed in beyond-Horndeski class theory \cite{Hirano:2018}. In any case, for a decisive detection of modified gravity signal, small-scale information will be key. In doing so, additional complication may arise from the post-Born correction (Eq.\ref{Eq:bisp:pb}), for which we must also rely on a non-linear prescription for the power spectrum in order to get an accurate prediction. For DGP model, a good approximation is to use the fitting formula in GR (i.e., HALOFIT)  and simply replace the linear power spectrum in GR with that in DGP. For $f(R)$ gravity, no such a simple treatment can work well, but there exists an extension of HALOFIT formula \cite{Zhao:2013dza}. Since such extensions are not readily available in general models, one may alternatively consider  the prediction based on perturbation theory  \cite{Taruya2016,Bose:2016qun}, although it would not capture accurately the small-scale behaviors.

Another interesting point is that in the DGP case (left plot of Fig.~\ref{fig:mg}) we find the fitting formula gives a significantly different prediction from the one-loop. This may hint at a lack of screening effect included in the fitting formula as only the second-order kernel is modified in this case. To investigate this further, one would require modified gravity simulation measurements and is beyond the scope of this work. 
 
\section{Summary} \label{sec:summary}

The CMB lensing bi-spectrum will be soon detected from upcoming CMB experiments, and will become an important observable, complementary to the lensing power spectrum. In this paper, making use of full-sky lensing simulations, we have tested the analytic predictions of the CMB lensing bi-spectrum based on existing fitting formulae of the dark matter bi-spectrum and on perturbation theory.  We found that the agreement between the simulation and fitting formulae depends on the configuration and scales of the bi-spectrum. In the equilateral case, the bi-spectrum obtained from the simulation is in good agreement with that derived from the fitting formulae. On the other hand, the squeezed bi-spectrum from the simulation deviates significantly from that predicted by the fitting formulae. 
{
We also found that a significant discrepancy also appears in the squeezed matter bi-spectrum at low $z$, indicating that the discrepancy in the squeezed lensing bi-spectrum probably comes mostly from the inconsistency between the analytic prediction and simulation results of the matter bi-spectrum. 
}
We have also compared the measured equilateral and squeezed configurations with one-loop perturbation theory. For the equilateral case we find that one-loop perturbation theory gives a much better prediction than tree-level calculation, but the scales of applicability is still restricted to rather low multipoles ($\ell \leq 600$). A notable point may be that the one-loop prediction becomes comparable to the fitting formula for the squeezed configuration up to $\ell \leq 1000$, though both fail to model the measured bi-spectrum as previously mentioned.  

Using one-loop perturbation theory, we have also discussed the effects of modified gravity on the lensing bi-spectrum. Considering DGP and $f(R)$ gravity models, we find that the signal of possible modification in the equilateral configuration is small at multipoles $\ell\lesssim800$, with a maximum of $\sim 1\%$ deviation from GR in DGP model and $\sim 4\%$ in $f(R)$ gravity. Such a signal cannot be distinguished even with stage-IV surveys, however, a large deviation from GR is still allowed in beyond-Horndeski class theory. Accurately modeling bi-spectrum in general theories of gravity is thus crucial especially at small scales. 

Further, we discussed the accuracy of the lensing bi-spectrum predictions while leaving aside possible practical issues in measuring the lensing bi-spectrum with real data. For example, with the quadratic estimator to reconstruct the lensing convergence, the measured bi-spectrum is a six-point correlation in CMB anisotropies. Similar to the lensing power spectrum measurement, an accurate subtraction of the disconnected six-point correlation is required. More practical issues on the lensing bi-spectrum measurement will be investigated elsewhere. 

Another interesting direction would be to revise the fitting formula of the matter bi-spectrum as we see the discrepancy between the simulated and analytic matter bi-spectrum in the squeezed configuration. The current fitting formula of the matter bi-spectrum is derived within a limited range of scale, configuration, and cosmological parameters. Since the bi-spectrum contains additional information on cosmology, an accurate fitting formula of the matter bi-spectrum will be necessary for a fast computation of the bi-spectrum prediction in future cosmological implications. The updated fitting formula will be also useful for studies on not only the CMB lensing but also, e.g., the galaxy weak lensing and galaxy clustering. We leave these for our future work.

\begin{acknowledgments}
TN acknowledges the support from the Ministry of Science and Technology (MOST), Taiwan, R.O.C. through the MOST research project grants (no. 107-2112-M-002-002-MY3). This research used resources of the National Energy Research Scientific Computing Center, which is supported by the Office of Science of the U.S. Department of Energy under Contract No. DE-AC02-05CH11231. BB acknowledges support from JSPS International Research Fellowship PE17043 and from the Swiss National Science Foundation (SNSF) Professorship grant No.170547. This work was supported in part by MEXT/JSPS KAKENHI Grant Number JP15H05899 (AT), JP16H03977 (AT), 15H05893 (RT) and 17H01131 (RT). 
Numerical computations were in part carried out on Cray XC50 at Center for Computational Astrophysics, National Astronomical Observatory of Japan.
\end{acknowledgments}

\onecolumngrid
\appendix

\section{Measurement of the bi-spectrum from full-sky simulation} \label{app:code}

Here we show the verification of our pipeline for measuring the bi-spectrum from the simulated CMB lensing convergence map. To test the algorithm of \eq{Eq:bispec:binned}, we consider the following non-Gaussian fluctuations: 
\al{
	a(\hatn) \equiv g(\hatn) + [g^2(\hatn)-\ave{g^2}] \,,
}
where $g(\hatn)$ is a random Gaussian map generated by a power spectrum, $C_\l$. 
We use the lensing convergence power spectrum to generate $g(\hatn)$. 
The harmonic coefficients at $\l>0$ are obtained as 
\al{
	a_{\l m} &= g_{\l m} + \Int{2}{\hatn'}{} Y^*_{\l m}(\hatn') g^2(\hatn') 
    	\notag \\
    	&= g_{\l m} + \Int{2}{\hatn'}{}\sum_{\l'\l''m'm''}
    		Y^*_{\l m}(\hatn')Y_{\l'm'}(\hatn')Y_{\l''m''}(\hatn')g_{\l'm'}g_{\l''m''}
        \notag \\
    	&= g_{\l m} + (-1)^m\sum_{\l'\l''m'm''}G^{\l\l'\l''}_{-m,m',m''}g_{\l'm'}g_{\l''m''}
	\,,
}
with $g_{\l m}$ being the harmonic coefficients of $g(\hatn)$. The Gaunt integral is defined as
\al{
	G^{\l_1\l_2\l_3}_{m_1m_2m_3} &\equiv 
    	\Int{2}{\hatn}{} Y_{\l_1m_1}(\hatn)Y_{\l_2m_2}(\hatn)Y_{\l_3m_3}(\hatn)
		\notag \\
		&= h_{\l_1\l_2\l_3}\Wjm{\l_1}{\l_2}{\l_3}{m_1}{m_2}{m_3} 
    \,, \label{Eq:gaunt}
}
where $h_{\l_1\l_2\l_3}$ is defined in \eq{Eq:hlll}. The Gaunt integral vanishes if $\l_1+\l_2+\l_3$ is an odd integer. Taking the complex conjugate of the above equation, we obtain $G^{\l_1\l_2\l_3}_{m_1m_2m_3}=(-1)^{m_1+m_2+m_3}G^{\l_1\l_2\l_3}_{-m_1,-m_2,-m_3}$. By definition, the Gaunt integral satisfies the following symmetric property:
\al{
	G^{\l_1\l_2\l_3}_{m_1m_2m_3} = G^{\l_2\l_3\l_1}_{m_2m_3m_1} = G^{\l_3\l_1\l_2}_{m_3m_1m_2}
    \,. 
}
Using the properties of the Wigner 3j symbol, 
\al{
	\sum_{m_1m_2m_3}[G^{\l_1\l_2\l_3}_{m_1m_2m_3}]^2 = h^2_{\l_1\l_2\l_3} \,. 
    \label{Eq:gaunt:sum}
}

The expected reduced bi-spectrum is decomposed into the contribution from the four and six point correlations. The contribution from the four point correlation is given by 
\al{
	b^{ggg^2}_{\l_1\l_2\l_3}
 	   &= h^{-2}_{\l_1\l_2\l_3} \sum_{m_i}G^{\l_1\l_2\l_3}_{m_1m_2m_3}
        \ave{a_{\l_1m_1}a_{\l_2m_2}a_{\l_3m_3}}
        \notag \\
 	   &= h^{-2}_{\l_1\l_2\l_3} \sum_{m_i}G^{\l_1\l_2\l_3}_{m_1m_2m_3}
        (-1)^{m_3}\sum_{\l'\l''m'm''}G^{\l_3\l'\l''}_{-m_3,m',m''}
        \ave{g_{\l_1m_1}g_{\l_2m_2}g_{\l'm'}g_{\l''m''}}
        + (\text{2 perms.}) 
    \,. 
}
Since $g_{\l m}$ is a random Gaussian fields, the four point correlation, $\ave{g_{\l_1m_1}g_{\l_2m_2}g_{\l'm'}g_{\l''m''}}$, is decomposed into the three terms by the Wick theorem. However, the term $\ave{g_{\l_1m_1}g_{\l_2m_2}}\ave{g_{\l'm'}g_{\l''m''}}$ provides the disconnected three-point correlation ($\ave{a_{\l_1m_1}a_{\l_2m_2}}\ave{a_{\l_3m_3}}$), and should be ignored in the above equation. Then we obtain
\al{
	b^{ggg^2}_{\l_1\l_2\l_3}
 	   &= h^{-2}_{\l_1\l_2\l_3} \sum_{m_i}G^{\l_1\l_2\l_3}_{m_1m_2m_3}
        (-1)^{m_3}\sum_{\l'\l''m'm''}G^{\l_3\l'\l''}_{-m_3,m',m''} 
       \notag \\
       &\quad \times (-1)^{m_1+m_2}C_{\l_1}C_{\l_2}
       [\delta_{\l_1\l'}\delta_{\l_2\l''}\delta_{m_1,-m'}\delta_{m_2,-m''}
    	+ \delta_{\l_1\l'}\delta_{\l_2\l''}\delta_{m_1,-m'}\delta_{m_2,-m''}]
        + (\text{2 perms.}) 
       \notag \\
 	   &= 2C_{\l_1}C_{\l_2}h^{-2}_{\l_1\l_2\l_3} \sum_{m_i}G^{\l_1\l_2\l_3}_{m_1m_2m_3}
        (-1)^{m_1+m_2+m_3}G^{\l_3\l_1\l_2}_{-m_3,-m_1,-m_2}
        + (\text{2 perms.}) 
       \notag \\
 	   &= 2[C_{\l_1}C_{\l_2}+(\text{2 perms.})]h^{-2}_{\l_1\l_2\l_3}
    \,. \label{Eq:bispec:test}
}
From the second to the third equation, we use \eq{Eq:gaunt:sum}. 

The contribution from the six point correlation contains the product of the three angular power spectrum. Thus, we choose the amplitude of the angular power spectrum so that the terms involving the six point correlation is negligible. 

\begin{figure*}
\bc
\includegraphics[width=89mm,clip]{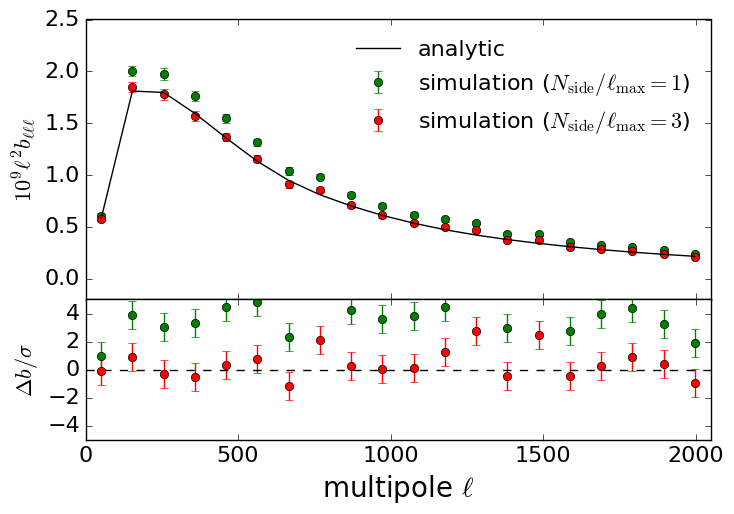}
\includegraphics[width=89mm,clip]{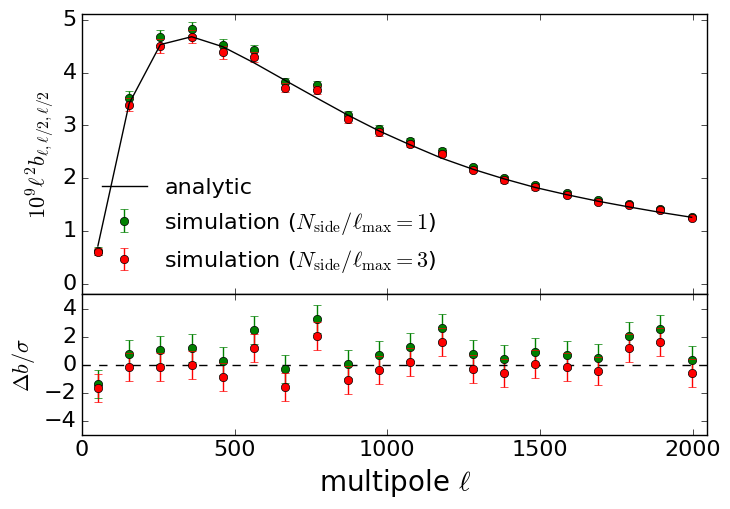}
\includegraphics[width=89mm,clip]{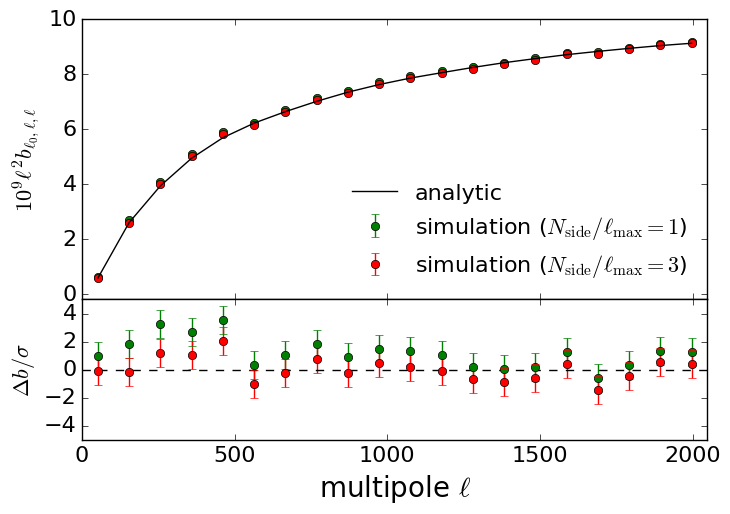}
\includegraphics[width=89mm,clip]{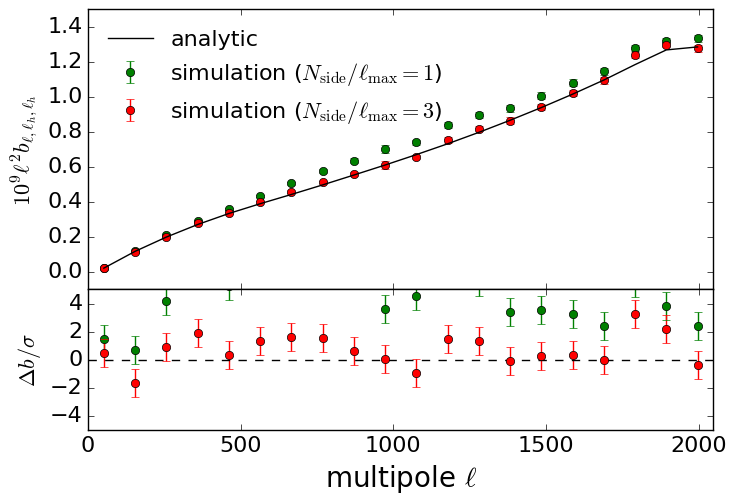}
\caption{
The reduced bi-spectrum obtained from the simple non-Gaussian simulation compares with the analytic formula. The binned analytic bi-spectrum is obtained from \eq{Eq:bispec:test}. The simulation results are shown for two cases with different values of the numerical parameter, $N_{\rm side}$ (see text). Note that in the lower panel we take the difference between the analytic and simulated bi-spectrum and then divide it with the $1\sigma$ simulation error. 
}
\label{fig:app:test}
\ec
\end{figure*}

Fig.~\ref{fig:app:test} shows the validity of our code for measuring the reduced bi-spectrum from simulations. The binned analytic bi-spectrum is obtained from \eq{Eq:bispec:binned} where the reduced bi-spectrum at each multipole is given by \eq{Eq:bispec:test}. In measuring the bi-spectrum from simulations, a resolution parameter, $N_{\rm side}$, in obtaining $\kappa_i(\hatn)$ and $h_i(\hatn)$ in Eqs.~\eqref{Eq:bispec:sim} and \eqref{Eq:bispec:norm} is chosen for each $i$ (multipole bin). We show the cases with $N_{\rm side}=\l^i_{\rm max}$ or $3\l^i_{\rm max}$ where $\l^i_{\rm max}$ is the maximum multipole at the $i$th bin. 

\section{Fitting formulas of the matter bi-spectrum} \label{app:fit}

Here we briefly summarize the fitting formulas of the matter bi-spectrum used in this paper. Ref.~\cite{Scoccimarro:2001} obtains the fitting formula of the matter bi-spectrum as 
\al{
	a(k,z) &= \frac{1+\{\sigma_8(z)\}^{a_6}\sqrt{0.7Q(n_{\rm eff})}(qa_1)^{n_{\rm eff}+a_2}}{1+(qa_1)^{n_{\rm eff}+a_2}} \\
	b(k,z) &= \frac{1+0.2a_3(n_{\rm eff}+3)(qa_7)^{n_{\rm eff}+3+a_8}}{1+(qa_7)^{n_{\rm eff}+3.5+a_8}} \\
	c(k,z) &= \frac{1+[4.5a_4/(1.5+(n_{\rm eff}+3)^4)](qa_5)^{n_{\rm eff}+3+a_9}}{1+(qa_5)^{n_{\rm eff}+3.5+a_9}}
	\,, 
}
with $Q(x)=(4-2^x)/(1+2^{x+1})$. Here, $q=k/k_{\rm NL}$ with $k_{\rm NL}$ where $4\pi k^3_{\rm NL}P_{\rm m}^{\rm lin}(k_{\rm NL})=1$. The quantity $\sigma_8(z)$ is the variance of the matter density fluctuations smoothed with a top-hat sphere of radius $8h^{-1}$Mpc at redshift $z$. $n_{\rm eff}\equiv d\ln P_{\rm m}^{\rm lin}(k)/d\ln k$, is the effective spectral index of the linear power spectrum, $P_{\rm m}^{\rm lin}(k)$. We apply the smoothing to compute $n_{\rm eff}$ in order to remove un-physical oscillation as discussed in Ref.~\cite{Gil-Marin:2012}. The parameters, $a_i$, are determined by fitting results of N-body simulations, which yields \cite{Scoccimarro:2001}
\al{
	a_1 &= 0.250 & a_2 &= 3.50 & a_3 &= 2.00 & a_4 &= 1.00 & a_5 &=2.00 & a_6 &= -0.200 &
    a_7 &= 1.00 & a_8 &= 0.00 & a_9 &= 0.00 \,.
    \label{Eq:SC}
}
Later on, Ref.~\cite{Gil-Marin:2012} proposed an improved fit given by  
\al{
	a_1 &= 0.484 & a_2 &= 3.74 & a_3 &= -0.849 & a_4 &= 0.392 & a_5 &= 1.01 & a_6 &= -0.575 &
    a_7 &= 0.128 & a_8 &= -0.722 & a_9 &= -0.926 \,.
    \label{Eq:GM}
}

\section{One-loop matter bi-spectrum} \label{app:spt}

The one-loop matter bi-spectrum is defined as 
\begin{align}
B^{1-{\rm loop}}(k_1,k_2,\theta;a) = &  B^{112}(k_1,k_2,\theta;a) \nonumber \\ & + [B^{222}(k_1,k_2,\theta;a)  +  B^{321}(k_1,k_2,\theta;a) + B^{114}(k_1,k_2,\theta;a)] \label{1loopbs},
\end{align}
where $\theta = \cos^{-1}{(\hat{\bfk}_1\cdot \hat{\bfk}_2)}$. The tree level ($B^{112}$) and one-loop terms (terms in square brackets) are defined in the usual way 
\begin{align}
\langle \delta_{n_1}(\bfk_1) \delta_{n_2}(\bfk_2) \delta_{n_3}(\bfk_3) \rangle &=
(2\pi)^3\delta_{\rm D}(\bfk_1+\bfk_2 + \bfk_3 )\,B^{n_1n_2n_3}(\bfk_1,\bfk_2,\bfk_3),
\end{align}
where we must add all permutations of the over-density perturbations, $\delta_n$, on the LHS, $n$ denoting the order of the perturbation. The angled brackets denote an ensemble average and under the assumptions of perturbation theory these averages can be decomposed into a product of linear power spectra convolved with perturbative kernels (see \cite{Bernardeau:2001qr} for a review). These kernels are determined by solving energy and momentum conservation equations order by order. We direct the reader to \cite{Bose:2018zpk} for a full description of this procedure for general models of gravity, including the nDGP and $f(R)$ models considered in this paper.

\section{Discrepancy in the squeezed lensing bi-spectrum} \label{app:squeez}

Here we discuss possible sources for the huge discrepancy which we found in the squeezed configuration between the theory based on fitting formulae and the direct simulation measurements 

\subsection{Role of non-linearity}

\begin{figure*}[t]
\bc
\includegraphics[width=178mm,clip]{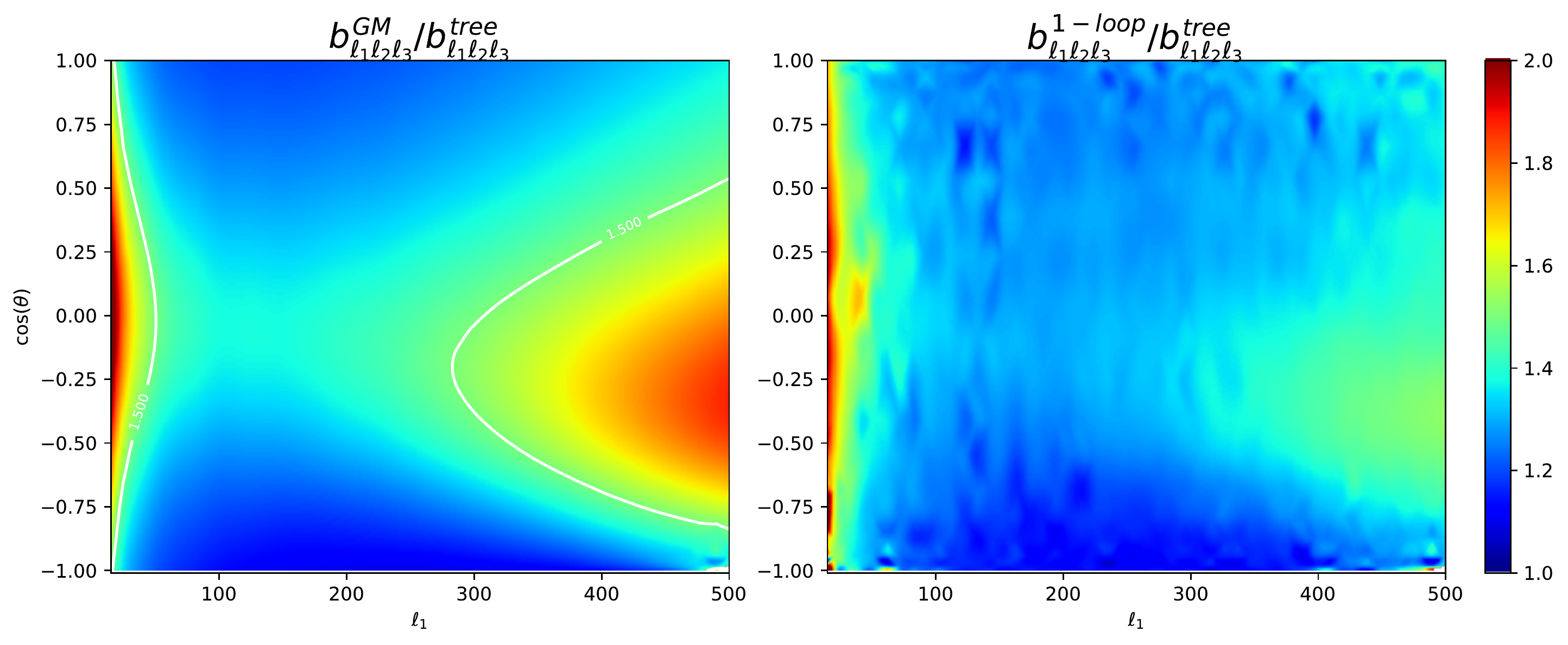}
\caption{
Contours showing the ratio of the non-linear prediction for the LSS contribution to the CMB lensed bi-spectrum (GM on left and 1-loop on right) to the tree level prediction. In both contours $\ell_2 =500$, with $\ell_3^2 = \ell_1^2 + \ell_2^2 - \ell_2 \ell_1 \cos(\theta)$ by the triangle condition.
}
\label{fig:bispec:contour}
\ec
\end{figure*}

The first check we perform is to see which configurations encode the most non-linear information. Those configurations with the most non-linearity will naturally be modelled the worst by the theoretical predictions. Fig.\ref{fig:bispec:contour} shows a contour of the ratio of the non-linear prediction to the tree level prediction for the LSS contribution to the CMB lensing bi-spectrum for various configurations. Specifically we show the GM (left) and 1-loop (right). We fix $\ell_2 = 500$ and vary $\ell_1$ and the angle between $\ell_1$ and $\ell_2$. The equilateral configuration has a good deal of non-linearity ($\ell_1 = 500, \cos(\theta) = -0.5$) which is expected as all modes are at their maximum multipole. We find in this case we have a $45\%$ enhancement and $33\%$ enhancement over the tree level prediction for GM and 1-loop respectively. In the squeezed limit we also find a lot of non-linear information ($\ell_1=50, \cos(\theta)=0.05$) with a $33\%$ and $30\%$ enhancement for GM and 
one-loop respectively. Further, in the very squeezed limit ($\ell_1 \rightarrow 10$) we get enhancements up to $60\%$. When binning, as the simulation measurement is, these highly non-linear configurations will be included and so failure in the theoretical predictions is likely to become an issue. Finally, we also note that the GM formula systematically predicts significantly more non-linearity than the one-loop for these limits which further suggests inaccuracy of the fit when  considering the bottom left plot of Fig.\ref{fig:bispec:b20}. 

Of course to perform a more accurate investigation one would ideally plot the same contour but with simulation measurements in the numerator of the ratio, but this would be very computationally expensive. Despite this, this does serve as a good indication as to which configurations are the most non-linear.

\subsection{Matter bi-spectrum}

\begin{figure*}[t]
\bc
\includegraphics[width=89mm,clip]{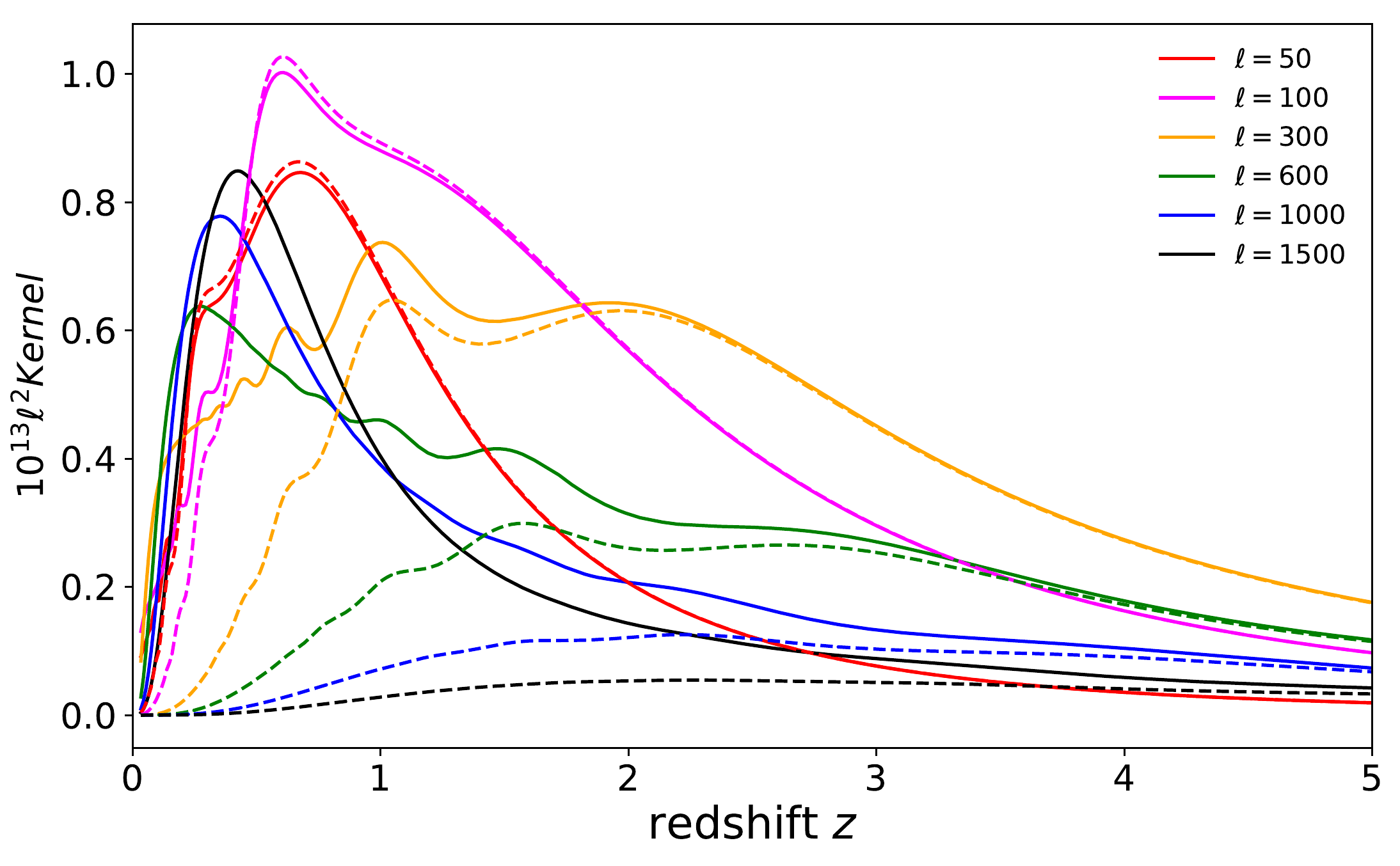}
\includegraphics[width=89mm,clip]{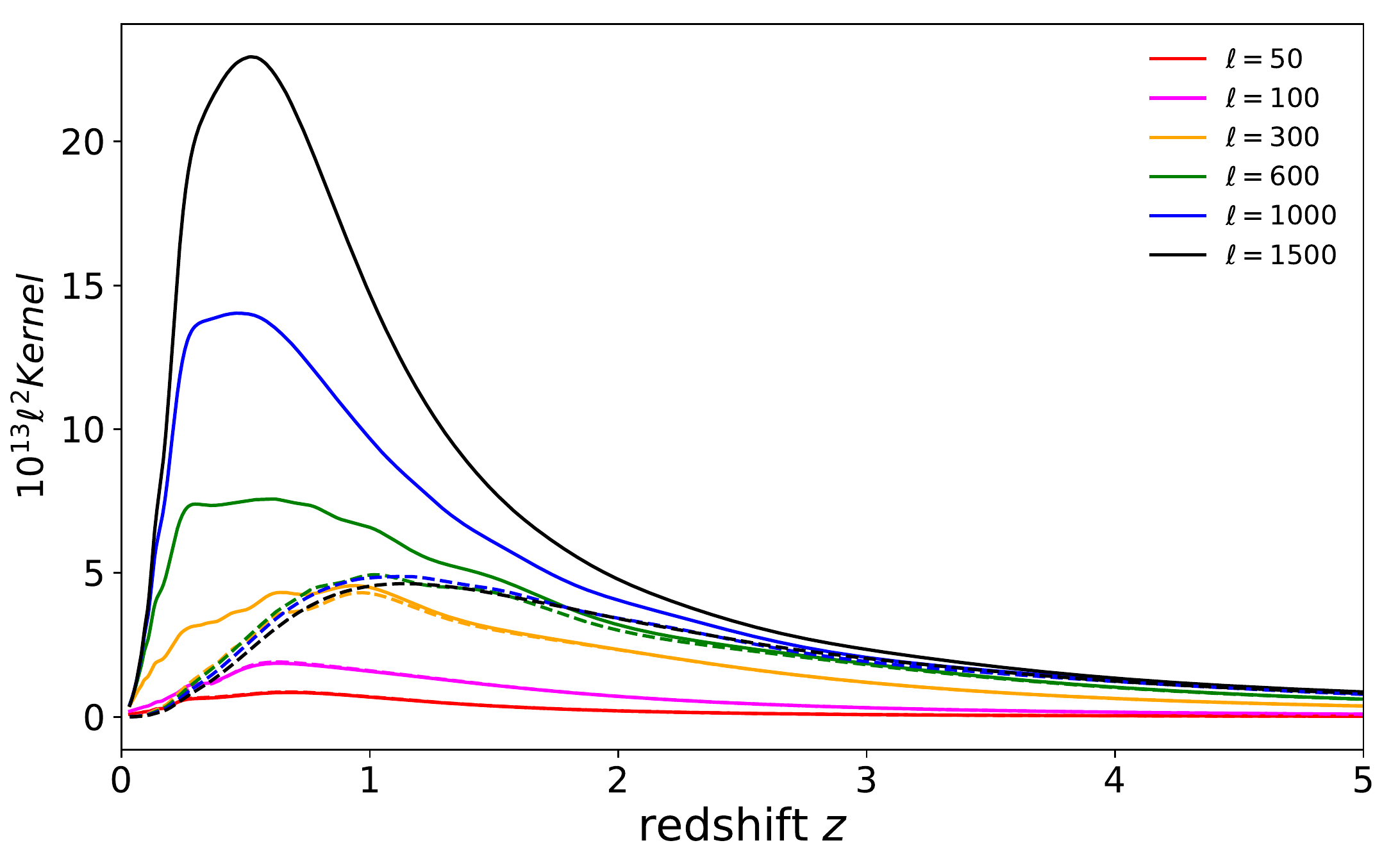}
\caption{
Equilateral (Left) and squeezed (Right) kernel of Eq.\ref{Eq:bisp:lss} as a function of redshift. Each curve denotes a different multipole $\ell = \ell_2 = \ell_3$. For the equilateral case $\ell_1 = \ell$ and for the squeezed case $\ell_1 = 50$. The solid lines show the GM fitting formula prediction while the dashed lines show the tree level prediction.
}
\label{fig:bispec:kernel}
\ec
\end{figure*}

\begin{figure*}[t]
\bc
\includegraphics[width=89mm,clip]{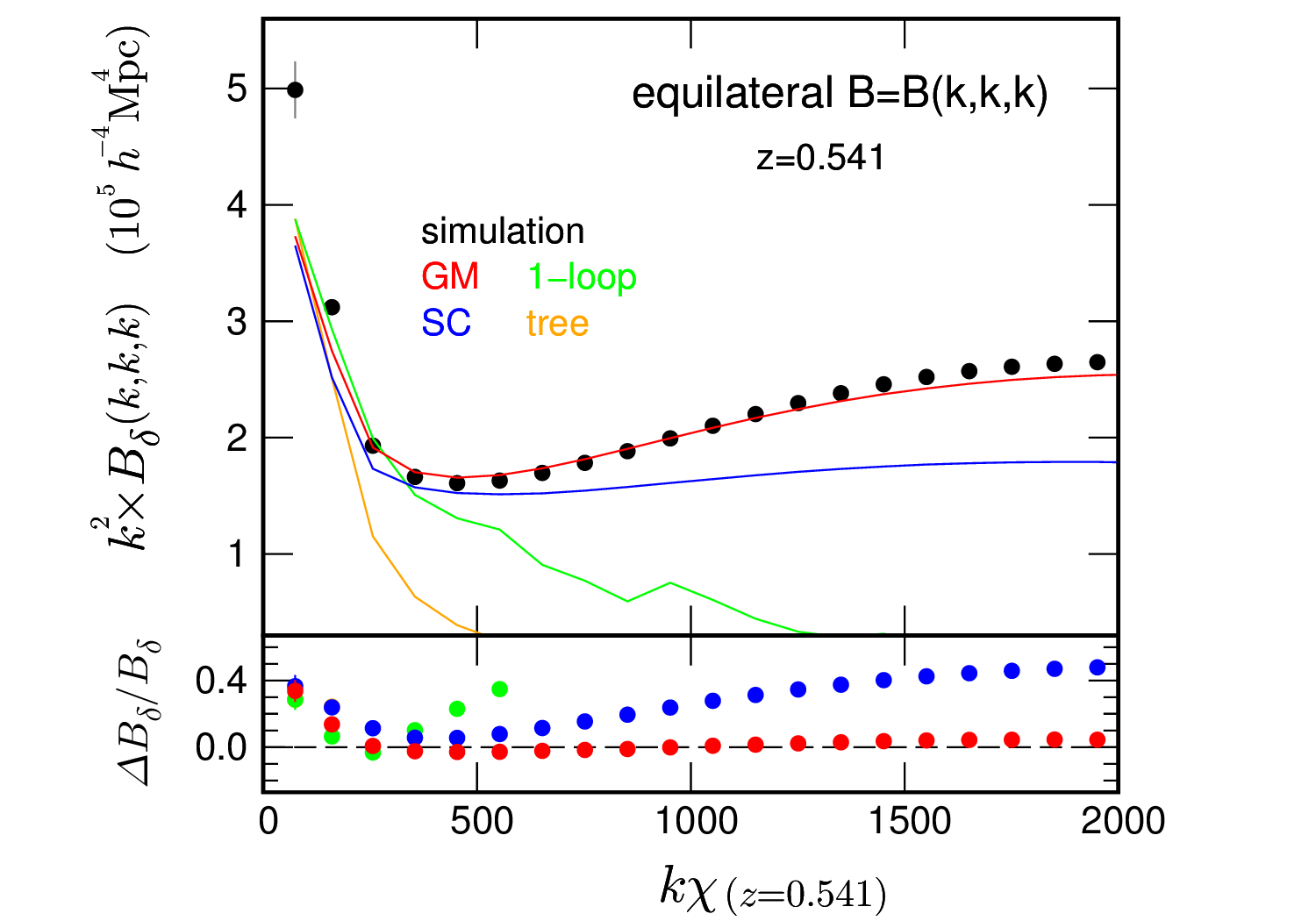}
\includegraphics[width=89mm,clip]{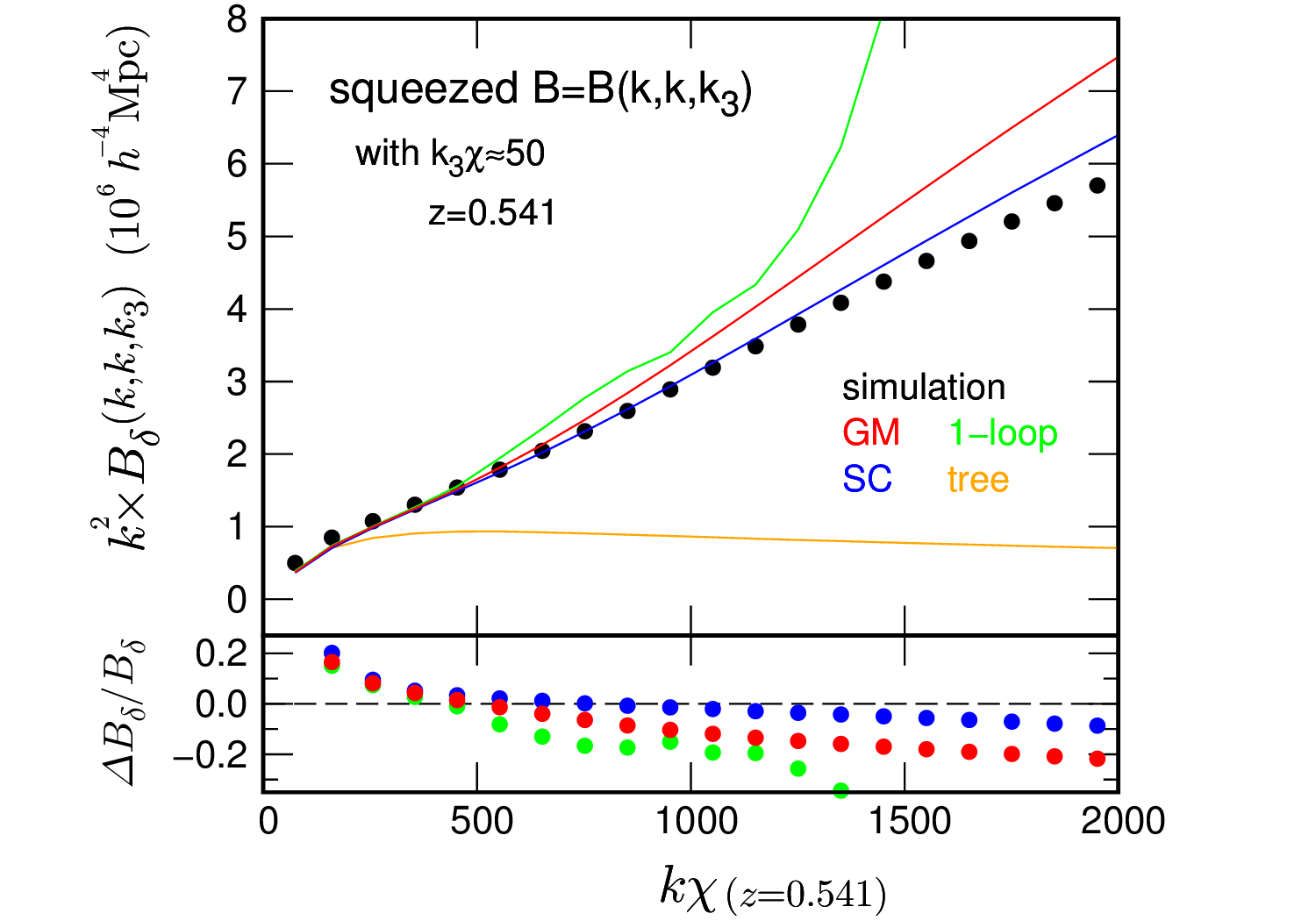}
\caption{
Matter bi-spectrum comparisons at $z=0.541$ for equilateral (left, $k_1=k_2=k_3=k$) and squeezed (right, $k_3 \chi \sim 50$, $k=k_1=k_2$) cases. The $x$-axis is the wavenumber $k$ times the comoving distance $\chi$, representing a multipole $\l$. The squeezed vector magnitude quoted is the weighted average of the smallest bin, $k_3 = 0.021\,h$\,Mpc$^{-1}$, where we take 20 linearly spaced bins from $1 \leq k_3 \chi \leq 2000$. The errors quoted in the figure are the variance over 6 realizations of the simulations.
}
\label{fig:bk_compare_z0.54}
\ec
\end{figure*}

\begin{figure*}[t]
\bc
\includegraphics[width=89mm,clip]{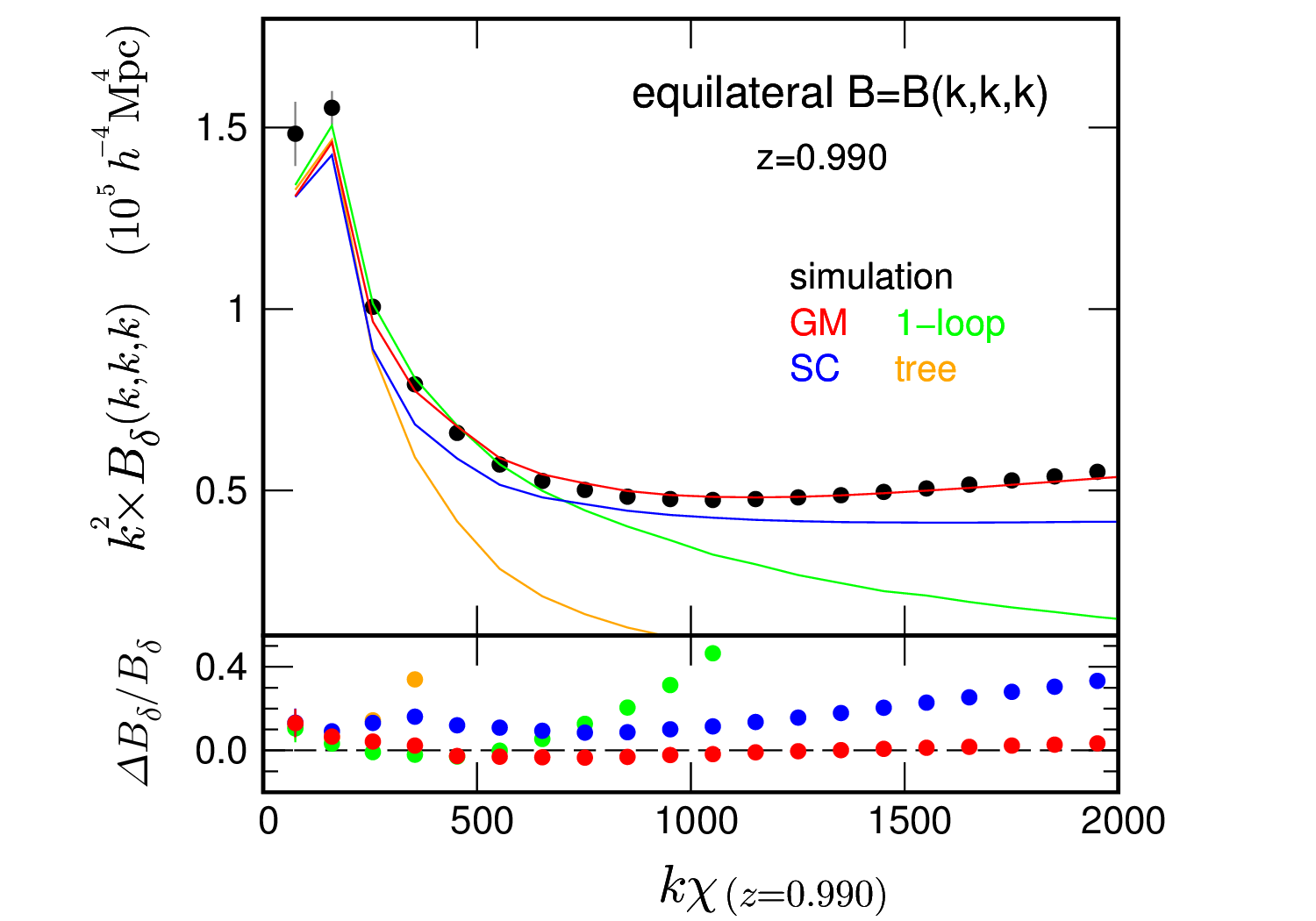}
\includegraphics[width=89mm,clip]{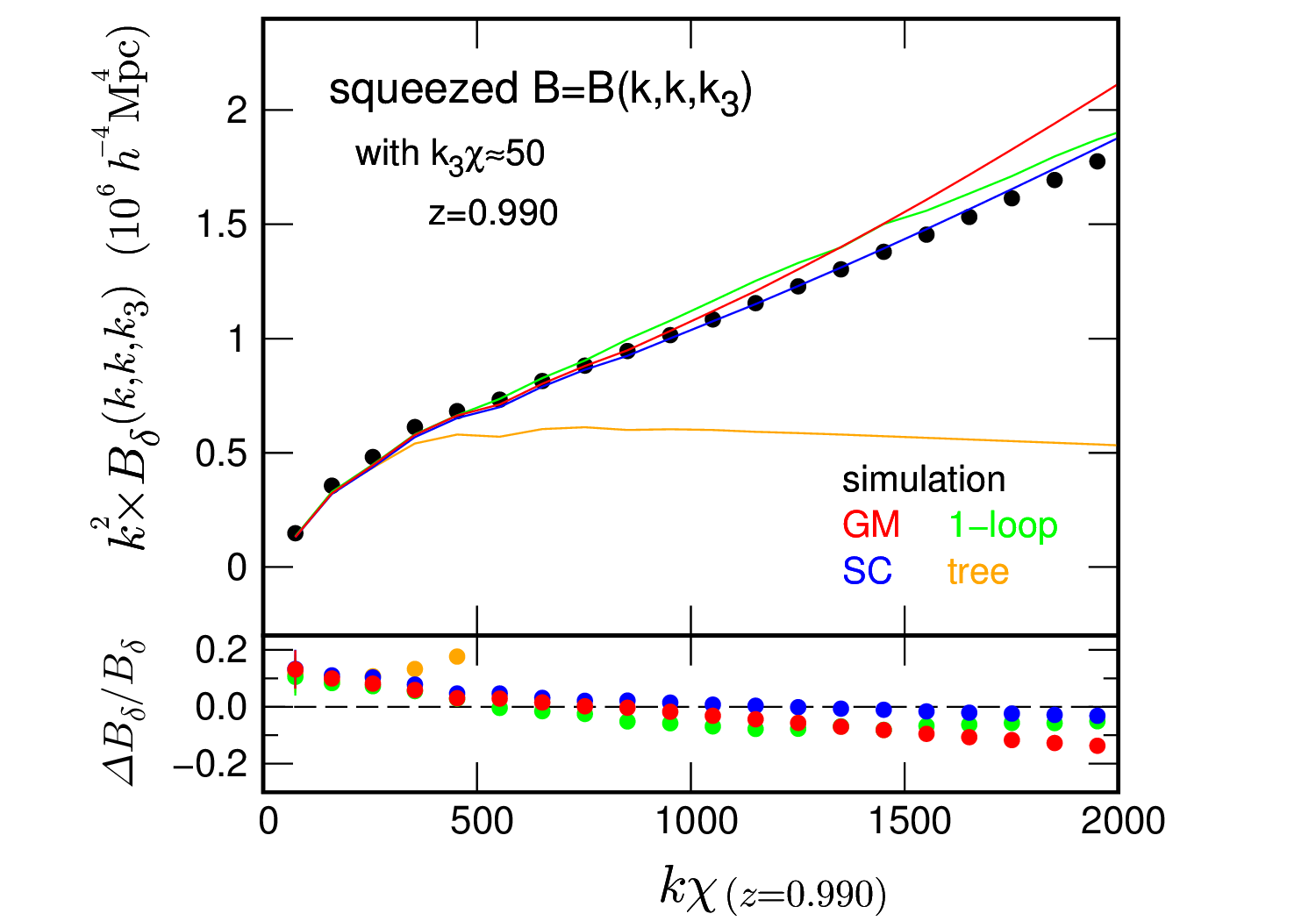}
\caption{
Same as Fig.\ref{fig:bk_compare_z0.54} but at $z=0.990$.
}
\label{fig:bk_compare_z0.99}
\ec
\end{figure*}

Given the good match of the post-Born prediction with simulation measurements seen in Fig.\ref{fig:bispec:postborn}), it is strongly suggested that the discrepancy comes from the LSS contribution, theoretically predicted by \eq{Eq:bisp:lss}. To investigate this issue we first have plotted the kernel of \eq{Eq:bisp:lss} as a function of redshift in Fig.\ref{fig:bispec:kernel} for the equilateral (left) and squeezed (right) configurations. Each curve denotes a different multipole with the dashed lines being the tree level prediction and the solid the GM prediction. We find that in the squeezed limit the kernel with the GM fitting formula peaks in the range $z\sim 0.5$, and for multipoles $\l\geq 600$ the peak is an order of magnitude larger than the equilateral shape. 
Given this we check the consistency of the squeezed matter bi-spectrum obtained from simulation and fitting formula at $z\sim 0.5$. Figs.~\ref{fig:bk_compare_z0.54} and \ref{fig:bk_compare_z0.99}  plot the matter bi-spectra at $z=0.541$ and $0.990$ measured from the N-body data used to construct the CMB maps. The box-sizes are $L=1800\,h^{-1}$Mpc (for $z=0.541$) and $2700\,h^{-1}$Mpc (for $z=0.990$) with $2048^3$ particles. The matter bi-spectrum at $z\sim0.5$ gives the most contribution to the CMB bi-spectrum from Fig.~\ref{fig:bispec:kernel}. The black circles with bars are mean with errors from $6$ realizations. The GM fitting formula gives a good agreement for equilateral case but slightly over predicts for the squeezed case at small scales. This discrepancy is seen even for half $k$-bin width. This trend seems consistent with the CMB lensing bi-spectrum. Such a discrepancy can lead to the significant over-prediction noted at small scales in Fig.~\ref{fig:bispec:b20} once we integrate over all redshifts. The discrepancy noted at large scales in Fig.~\ref{fig:bispec:b20} could be a combination of effects including post-Born effects. 

As mentioned in the main text this discrepancy may be ascribed to an inaccuracy in the GM fitting formula. Indeed, the formula is not fit to very squeezed configurations and was only fit in the range of $0.03\,h$\,Mpc$^{-1}$$ \leq k \leq 0.4\,h$\,Mpc$^{-1}$ \cite{Gil-Marin:2012}, corresponding to $46 \leq \ell \leq 570$ at $z=0.541$. Further, the squeezed limit considered here for $100 \leq \ell$ has $ 0.9 \leq \theta /\pi$, $\theta$ being the angle between the `non-squeezed' wave vectors, while the formula is only fit for $\theta /\pi \leq 0.9$. Despite this possibility, we perform a number of further tests to clarify the issue. 

\subsection{Resolution of simulation}

As discussed in Ref.~\cite{Coulton:2018ebd}, the resolution of the simulation map also changes the small scale behavior of the squeezed lensing bi-spectrum. In the main text, we show the measurements of the bi-spectrum from the simulation of $N_{\rm side}=4096$. Here we check how the results depend on the simulation map resolution, $N_{\rm side}$. 

\begin{figure*}
\bc
\includegraphics[width=89mm,clip]{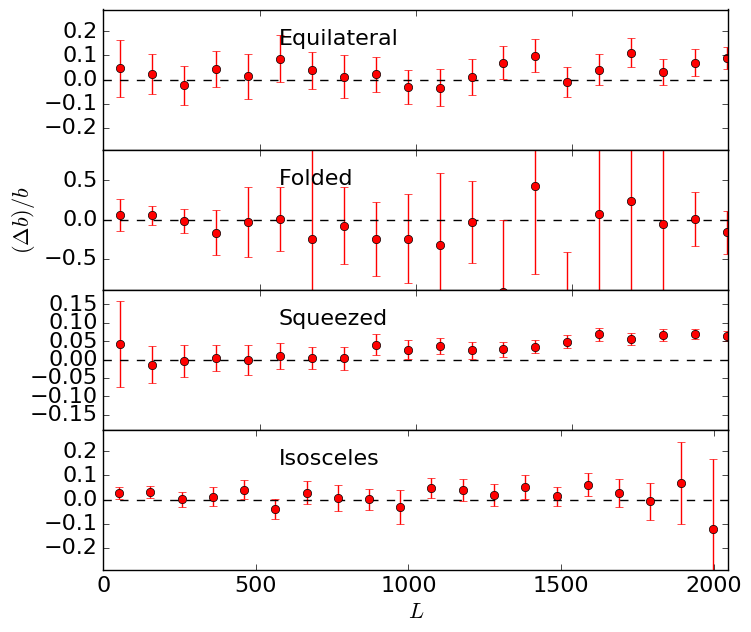}
\caption{
Difference of the bi-spectrum measured from simulations with different resolution. We plot the fractional difference, $\Delta b/b\equiv b^{N_{\rm side}=8192}/b^{N_{\rm side}=4096} - 1$ where $b^{N_{\rm side}=4096}$ and $b^{N_{\rm side}=8192}$ are measured bi-spectra obtained from the simulations of $N_{\rm side}=4096$ and $8192$, respectively. 
}
\label{fig:nres}
\ec
\end{figure*}

Fig.~\ref{fig:nres} shows the fractional difference of the bi-spectrum at each configuration, $\Delta b/b$, where $\Delta b/b\equiv b^{N_{\rm side}=8192}/b^{N_{\rm side}=4096} - 1$ is the fractional difference between the bi-spectra measured from the simulation of $N_{\rm side}=8192$ and $N_{\rm side}=4096$. In the squeezed case, the increase of the resolution enhances the amplitude of the bi-spectrum by $5-10\%$ at smaller scales. This increase is, however, too small compared to the discrepancy we found.

\subsection{Other possibilities}

\begin{figure}
\bc
\includegraphics[width=89mm,clip]{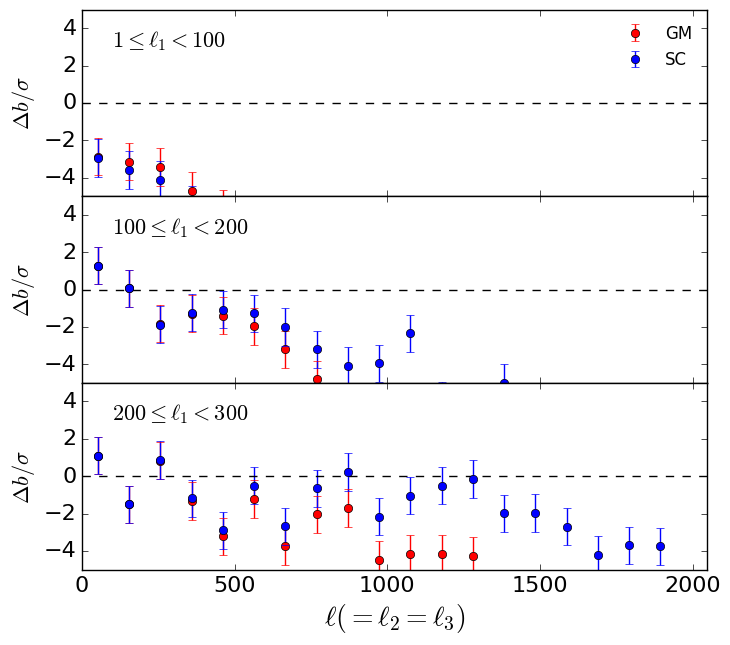}
\caption{
Squeezed bi-spectrum, $b_{\l_1\l\l}$, with varying the multipole bin for $\l_1$. 
}
\label{fig:bispec:l1dep}
\ec
\end{figure}

Fig.~\ref{fig:bispec:l1dep} shows the dependence of the discrepancy in the squeezed bi-spectrum on the multipole bin for $\l_1$. As the multipole range of $\l_1$ goes to smaller scales, the discrepancy between the analytic and simulated bi-spectrum is reduced. However, the analytic prediction always overestimates the simulation. Indeed, the discrepancy is still large when we choose a larger multipole bin to $\l_1$ which is much less effected by sample variance or possible large scale systematics. 

\begin{figure}
\bc
\includegraphics[width=89mm,clip]{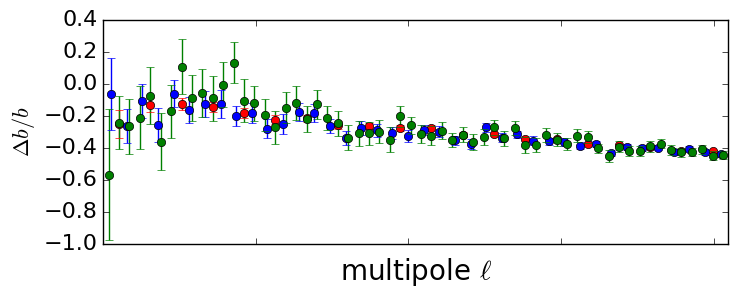}
\caption{
Fractional difference of squeezed bi-spectrum, $\Delta b/b$, with varying the number of multipole bins (red: 20 bins, blue: 40 bins, green: 60 bins). 
}
\label{fig:bispec:binnumber}
\ec
\end{figure}

Fig.~\ref{fig:bispec:binnumber} plots the fractional difference of the bi-spectrum with varying the number of multipole bins ($20,40$ and $60$). The discrepancy is almost independent from the number of multipole bins. 


\twocolumngrid
\bibliographystyle{mybst}
\bibliography{exp,main}

\end{document}